\documentclass[preprint]{aastex}

\newcommand{\be}{\begin{equation}}
\newcommand{\ee}{\end{equation}}

\newcommand{\discrim}{{ {\cal D} }}

\newcommand{\delx}{{ \delta \xi }}
\newcommand{\mchat}{{ { m}_0 }} 
\newcommand{\lamprime}{{ \Lambda^\prime_\ast }}
 
\newcommand{\mbench}{{ M_{\rm bench} }} 

\begin{document} 

\title{\bf AMBIPOLAR DIFFUSION IN MOLECULAR CLOUD CORES \\
AND THE GRAVOMAGNETO CATASTROPHE} 

\author{Fred C. Adams\altaffilmark{1,2} and Frank H. Shu\altaffilmark{3}} 

\altaffiltext{1}{Michigan Center for Theoretical Physics \\
Physics Department, University of Michigan, Ann Arbor, MI 48109} 

\altaffiltext{2}{Astronomy Department, University of Michigan, Ann Arbor, MI 48109} 

\altaffiltext{3}{Center for Astrophysics and Space Sciences \\  
Physics Department, University of California, La Jolla, CA 92093} 

\begin{abstract} 

This paper re-examines the problem of ambipolar diffusion as a
mechanism for the production and runaway evolution of centrally
condensed molecular cloud cores, a process that has been termed the
gravomagneto catastrophe.  Our calculation applies in the geometric
limit of a highly flattened core and allows for a semi-analytic
treatment of the full problem, although physical fixes are required to
resolve a poor representation of the central region. A noteworthy
feature of the overall formulation is that the solutions for the
ambipolar diffusion portion of the evolution for negative times 
($t < 0$) match smoothly onto the collapse solutions for positive
times ($t > 0$). The treatment shows that the resulting cores display
non-zero, but sub-magnetosonic, inward velocities at the end of the
diffusion epoch, in agreement with current observations. Another
important result is the derivation of an analytic relationship between
the dimensionless mass to flux ratio $\lambda_0\equiv f_0^{-1}$ of the
central regions produced by runaway core condensation and the
dimensionless measure of the rate of ambipolar diffusion $\epsilon$. 
In conjunction with previous work showing that ambipolar diffusion
takes place more quickly in the presence of turbulent fluctuations,
i.e., that the effective value of $\epsilon$ can be enhanced by
turbulence, the resultant theory provides a viable working hypothesis
for the formation of isolated molecular-cloud cores and their
subsequent collapse to form stars and planetary systems.

\end{abstract} 


\section{INTRODUCTION} 

Since molecular clouds are supported, at least in part, by magnetic
fields, the removal of magnetic fields represents an important
component of the star formation process. In the most studied scenario,
field removal occurs through the action of ambipolar diffusion,
wherein magnetic fields are tied to the ionized component, which
drifts relative to the more dominant neutral component of the gas
(Mestel \& Spitzer 1956, Mouschovias 1976, Nakano 1979, Shu 1983,
Nakano 1984, Lizano \& Shu 1989, Basu \& Mouschovias 1994).  The
neutral gas thus condenses inward toward a quasi-hydrostatic state,
although perfect equilibrium is generally not reached. When the
condensing gas becomes sufficiently centrally concentrated, the
innermost regions of the structure begin to collapse dynamically onto
a growing point-like protostar and eventually approach ballistic
(free-fall) conditions.

Because of its large dynamic range in space and time, the process is
not easy to follow numerically.  The magnetic field and the forces
that it exerts are vector quantities, so the relevant diffusion and
dynamic equations are generally nonlinear, coupled partial
differential equations in multiple spatial dimensions. The assumption
of axial symmetry provides some simplification, but one still needs to
deal with two spatial dimensions and time.  In spite of these
complications, the net result is physically simple: As magnetic fields
diffuse outward, gas condenses inward to form a centrally concentrated
structure that approaches pure power-law distributions of gas density,
magnetic field, and other quantities (Nakano 1979, Lizano \& Shu 1989,
Basu \& Mouschovias 1994).  The goal of this paper is to demonstrate
in a simple mathematical fashion how the asymptotic state is reached
through nearly self-similar evolution toward a gravomagneto
catastrophe, wherein an infinite central concentration is formally
reached in finite time.  After this point of catastrophe, set equal to
the pivotal time $t=0$, the system experiences true dynamical collapse
to form a point-like star, which will be surrounded by a centrifugally
supported disk due to the pre-collapse rotation of the cloud core. 
Note that this study does not include the effects of rotation, so that
the collapse solutions found herein represent the outer portion of the
collapse flow; these solutions can then be matched onto inner solutions
that include rotation and other effects (e.g., Cassen \& Moosman 1981, 
Terebey et al. 1984, Jijina \& Adams 1996). 

Although molecular cloud cores experiencing ambipolar diffusion were
identified as playing a dominant role in the formation of isolated
low-mass stars two decades ago (Shu et al. 1987), recent observations
indicate that ambipolar diffusion takes place more rapidly than the
simple laminar description (e.g., Jijina et al.  1999). In addition,
nonzero inflow velocities are often observed in starless cores (Lee et
al. 2001, Harvey et al. 2002), which are assumed to be pre-collapse
states.  Contrary to popular perception, both of these properties can
be accommodated within the picture of core formation via ambipolar
diffusion. Indeed, one of the principal results of this paper is to
calculate the nonzero, but sub-magnetosonic, inward velocity resulting
from the ambipolar diffusion process. Despite this faster evolution,
molecular cloud cores are well-defined entities and not transient,
turbulent phenomena (Lada et al. 2007).  As a consequence of their
linkage to strong magnetic fields, probably, these cores are also
restrained from moving ballisticly through their parent clouds
(Walsh et al. 2004).

The rest of this paper is organized as follows.  We present the
formulation of the problem in terms of physical variables in \S 2,
where we enforce axial symmetry and use a flattened approximation. The
following section (\S 3) outlines our approach to solving the
resulting problem.  We first apply a similarity transformation, which
converts the partial integro-differential equations into ordinary
integro-differential equations. Since the diffusion time scale is
comparable to or longer than the magnetosonic time scale needed to
cross from the core center to the boundary where the core attaches to
a common envelope, the solution to the induction equation itself
requires a more complicated approach.  In \S 4, we solve the
zeoreth-order condensation problem for $t < 0$ to describe the
approach to the pivotal instant $t = 0$ of gravomagneto catastrophe.
As a simplification, we use the monopole (split monopole)
approximation for the gravitational (magnetic tension) forces, thereby
transforming the integro-differential equations to ordinary
differential equations that are solved by standard methods. In \S 5,
we point out the shortcomings at large and small radii of the monopole
approximation, and we generalize the approach by adopting various
mathematical and physical fixes that show how the general physical
problem possesses condensation solutions that connect smoothly to
marginally critical envelopes.  A central result of this section is
the analytic derivation of a relationship between the dimensionless
rate of ambipolar diffusion $\epsilon$ and the dimensionless flux to
mass ratio $f_0 \equiv \lambda_0^{-1}$ of the central regions of the
condensing core at the moment of gravomagneto catastrophe.  In \S 6,
we demonstrate how the runaway condensation that characterizes
gravomagneto catastrophe transitions smoothly for $t > 0$ to
dynamically collapsing states that correspond to cores with accreting
point-like protostars, i.e., the infall-collapse solutions that have
been used widely in previous studies of star formation.  In \S 7, we
present a specific dimensional example to illustrate the typical
astronomical characteristics of the entire process on both sides of
the pivotal instant $t = 0$. We conclude in \S 8 with a summary of the
astronomical implications of our results.  Finally, in a series of
Appendices A through F, we develop and extend various technical points
encountered in the discussion of the text.

\section{FORMULATION} 

The basic evolutionary equations for a flattened, self-gravitating,
cloud core of surface density $\Sigma$ and radial velocity $u$,
threaded by a magnetic field with vertical component $B_z$, are taken
from the analysis of Shu \& Li (1997) to be as follows.  The equation
of continuity is given by
\be
{\partial \Sigma \over \partial t} + {1 \over \varpi} 
{\partial \over \partial \varpi} \left[ \varpi u \Sigma \right] 
= 0 \, .
\label{continuityeq}
\ee
The force equation is
\be 
{\partial u \over \partial t} + 
u {\partial u \over \partial \varpi} + {a^2\over \Sigma} 
{\partial \over \partial \varpi}(\Theta \Sigma) = g + \ell \, , 
\label{force}
\ee
where the acceleration produced by self-gravitation plus magnetic tension, 
$g + \ell$, is given by 
\be
g + \ell = {1 \over \varpi^2 \Sigma} \int_0^\infty 
K_0 (r/\varpi) \left[ - G \Sigma(\varpi) \Sigma(r) + 
{B_z(\varpi) B_z(r) \over (2 \pi)^2} \right] 2 \pi r dr \, , 
\ee
and the kernel $K_0$ is defined via 
\be
K_0 (q) = {1 \over 2 \pi} \int_0^{2 \pi}  
{(1 - q \cos \varphi) d \varphi \over 
(1 + q^2 - 2 q \cos \varphi)^{3/2} } \, . 
\label{kernelzero}
\ee 
In equation (\ref{force}), $a$ is the gaseous isothermal sound speed,
and $\Theta$ provides the correction for the effects of the magnetic
pressure (see Appendix A).  Finally, the induction equation, which
governs the evolution of the vertical component of the magnetic field
threading the core in the presence of ambipolar diffusion, takes the
form (see Appendix B)
\be 
\Sigma \left\{ 
{\partial B_z \over \partial t} + {1 \over \varpi} 
{\partial \over \partial \varpi} \left[ \varpi u B_z \right] 
\right\} = {1 \over \varpi} {\partial \over \partial \varpi} 
\left[ { (2 z_0)^{1/2} \over 2 \pi \gamma {\cal C} } 
{\varpi B_z^2 B_\varpi^+ \over \Sigma^{1/2} } \right] \, ,
\label{inductioneq} 
\ee
where we have defined the radial component of the field at the upper 
vertical surface of the core by
\be 
B_\varpi^+ = {1 \over \varpi^2} 
\int_0^\infty K_0 (r/\varpi) B_z(r) \, r dr \, . 
\label{bplusdef} 
\ee 
The half-height $z_0$ appearing in equation (\ref{inductioneq}) is
defined by the assumed vertical hydrostatic equilibrium (Appendix A).
The quantities $\gamma$ and $1/{\cal C}$ are, respectively, the usual
drag coefficient between ions and neutrals and the height-averaged
reciprocal coefficient for the ion mass-abundance (see Appendix B and
Chapter 27 of Shu 1992).  An attractive feature of the approach
presented in this paper is that we can delay specifying the actual
numerical value of the product $\gamma {\cal C}$ until it comes to
specifying dimensional scalings appropriate to specific astronomical
objects, as long as the combination of parameters given by equation
(\ref{epsdef}) below is a small number compared to unity.

We define the dimensionless ratio $\lambda$ of mass per unit area 
to flux per unit area according to 
\be
\lambda = {2 \pi G^{1/2} \Sigma \over B_z } \, . 
\label{lambdadef}
\ee 
Appendix A derives expressions for $\Theta$ and $z_0$ in terms of
$\lambda$ for a magnetized singular isothermal disk, the form that our
inner core approaches asymptotically at the moment of gravomagneto
catastrophe. These relationships have the elegance of simplicity, and 
we adopt the approximation that the following expressions from
Appendix A hold for all time, i.e., 
\be 
\Theta = {2+\lambda^2 \over 1+\lambda^2} \qquad {\rm and} \qquad 
z_0 = \left( {\lambda^2 \over 1+\lambda^2}\right)
{a^2\over \pi G \Sigma}.
\label{closure}
\ee
Combined with equations (\ref{continuityeq}), (\ref{force}),
(\ref{inductioneq}), the relationships (\ref{lambdadef}) and
(\ref{closure}) give us a closed set of equations to solve for
$\Sigma$, $u$, $B_z$, $\lambda$, $\Theta$, and $z_0$.
 
\section{HOMOLOGY, SELF-SIMILARITY, AND ASYMPTOTICS} 

In this section, we construct a similarity transformation to recast
the problem in simpler form. First, we want to simplify the magnetic
induction equation by using equation (\ref{lambdadef}) and by making
the (usual) approximation that the combination $\gamma {\cal C}$ is a
constant during the phase of molecular-cloud core formation. As a
result, equation (\ref{inductioneq}) takes the form
\be
\Sigma^2 \left( {\partial \over \partial t} + 
u {\partial \over \partial \varpi} \right) {1 \over \lambda} 
= \sqrt{2\over \pi} \left( {a \over \gamma {\cal C}}\right) 
{1\over \varpi} {\partial \over \partial \varpi}  
\left[ { \varpi \Sigma B_\varpi^+ 
\over \lambda (1+\lambda^2)^{1/2} } \right] \, , 
\ee 
With this transformation, the definition of $B_\varpi^+$ 
becomes
\be
B_\varpi^+ = {2 \pi G^{1/2} \over \varpi^2} \int_0^\infty 
r \, dr \, K_0 (r/\varpi) {\Sigma(r) \over \lambda (r) } \, . 
\ee
Similarly, the force terms now have the form 
\be
g + \ell = {2 \pi G \over \varpi^2}  \int_0^\infty 
K_0 (r/\varpi) \Sigma(r) r \, dr \, \left[ - 1 + 
{1 \over \lambda(r) \lambda(\varpi) } \right] \, . 
\ee

\subsection{Basic Similarity Transformation} 

With the form of the magnetic induction equation specified, we now
transform from a $(\varpi, t)$ description to a $(x,t)$ description,
where we use the relations 
\be 
x = {\varpi \over a |t| } \, , \quad 
\Sigma(\varpi, t) = {a \over 2\pi G |t| } \tilde \sigma(x,t) \, , 
\quad u(\varpi, t) = a \tilde v(x,t) \, , \quad 
B_z(\varpi, t) = {a \over G^{1/2} |t|}\tilde \beta (x,t) \, . 
\label{dimensionless}
\ee 
In the simplest type of transformation, self-similarity of the first
kind (Barenblatt 1996), the functions $\tilde \sigma$, $\tilde v$, and
$\tilde \beta$ introduced here would be functions of the similarity
variable $x$ only. In this case, however, we allow the functions to
retain an additional time dependence to account for the fact
that amibipolar diffusion occurs on a longer time scale than the
runaway dynamics.  Notice also that we have written the time variable
in the coefficients with absolute value signs. We wish to mark the
pivotal time $t=0$ as the moment of gravomagneto catastrophe, so that
positive times correspond to the self-similar solutions of
gravitational collapse onto a point-like protostar (Li \& Shu 1997),
whereas negative times correspond to the epoch of ambipolar diffusion
in starless cores.  The start of the ambipolar diffusion process thus
corresponds to the limit $t \to -\infty$, and the end of the ambipolar 
diffusion epoch corresponds to the limit $t \to 0^-$. 

With this choice of transformation, the dimensionless 
mass-to-flux ratio is given by 
\be 
\lambda = \lambda (x,t) = {\tilde \sigma (x,t) \over \tilde \beta (x,t) } \, .
\ee 
We also define its inverse, i.e., the dimensionless flux-to-mass ratio,
\be 
f(x,t) \equiv {1 \over \lambda (x,t) } \, = 
{\tilde \beta (x,t) \over \tilde \sigma (x,t)} \, . 
\ee 
The derivatives then take the forms 
\be 
\Biggl( {\partial \over \partial t} \Biggr)_\varpi = 
\Biggl( {\partial \over \partial t} \Biggr)_x + 
{x \over |t|} {\partial \over \partial x} \, , 
\ee
and 
\be
\Biggl( {\partial \over \partial \varpi} \Biggr)_t = 
{1 \over a |t|} {\partial \over \partial x} \, .
\ee 
 
With this formulation, the equation of continuity is given by
\be 
|t|{\partial \tilde \sigma \over \partial t}+ \left( 1 + 
x {\partial \over \partial x} \right) \tilde \sigma + 
{1 \over x} {\partial \over \partial x} 
\left( x\tilde v\tilde \sigma \right) = 0 \, . 
\ee 
The force equation then becomes 
\be 
|t|{\partial {\tilde v} \over \partial t} + 
\left( x +\tilde v\right) {\partial\tilde v\over \partial x} 
+ {1 \over \tilde \sigma} {\partial \over \partial x}(\Theta \tilde \sigma) = 
\int_0^\infty K_0 \left( {y\over x}\right)  
\tilde \sigma (y,t) \left[ f(x,t) f(y,t)-1 \right] \, {y dy \over x^2} \,. 
\ee 
The induction equation can be written 
\be
\tilde \sigma^2\left[ |t|{\partial f\over \partial t} + 
(x + \tilde v) {\partial f \over \partial x}\right] 
= {\epsilon \over x} {\partial \over \partial x} \left[  x 
\tilde \sigma \tilde b \left( {f^2 \over \sqrt{1+f^2} }\right) \right] \, ,
\label{indeq}
\ee
where we have defined 
\be
\epsilon \equiv {\sqrt{8\pi G}\over \gamma {\cal C}} \, ,
\label{epsdef}
\ee
so that $\epsilon$ is a small dimensionless parameter of the problem
(essentially the ratio of dynamical time to the diffusion time; see
also Galli \& Shu 1993, who denoted a similar inverse ratio as a large
parameter $\chi$). In addition, the reduced radial magnetic field
$\tilde b$ is defined in terms of the integral
\be 
\tilde b(x,t) = {1 \over x^2} \int_0^\infty K_0 \left({y\over x} 
\right) \tilde \sigma(y,t) f(y,t) \, ydy \, . 
\label{bdef} 
\ee

If we use the traditional microscopic values of $\gamma = 3.5\times
10^{13}$ cm$^3$ g$^{-1}$ s$^{-1}$ and $\cal C$ = $2.0\times 10^{-16}$
cm$^{-3/2}$ g$^{1/2}$ (see Appendix B), we obtain $\epsilon \approx
0.18$.  This small, but not very small, value of $\epsilon$ allows for
an illuminating, but not highly accurate, attack on the problem of
molecular-cloud core formation and collapse.  In practice, turbulence
within the forming core may increase the effective diffusion
coefficient by a factor of a several (Zweibel 2002, Fatuzzo \& Adams
2002, Heitsch et al. 2004, Nakamura \& Li 2005), which makes
$\epsilon$ a marginally small parameter. On the other hand, if cosmic
ray fluxes are enhanced within molecular clouds (Fatuzzo et al. 2006),
the value of $\epsilon$ could be reduced by a factor of several. Thus,
we anticipate that $\epsilon$ might have large variations within
molecular clouds, accounting in part for the wide range of observed
core masses.  The formal theory developed here allows a
semi-analytical description of only those cores that form from regions
where $\epsilon \ll 1$, but we anticipate that many of the physical
insights gained from the formal analysis may carry over to the more
general case even when $\epsilon \sim 1$.

\subsection{Solution by Iteration}

From many numerical simulations, we know that the effect of ambipolar
diffusion in cloud cores is to try to redistribute the magnetic flux
from the inner region, where the flux-to-mass ratio $f$ has a
relatively low, constant, value to the outer region where $f$ has a
relatively high, constant, value.  By relatively high, we mean
typically $f\approx 1$; and by relatively low, we mean typically $f
\approx 1/2$.  Thus, $f$ may vary only by a factor of 2 over a dynamic
range in spatial scale of $10^4$, say, from $10^{-4}$ pc to 1 pc,
whereas the volume density $\propto \Sigma/2z_0$ over the same range
of radii might differ by a factor of as much as $10^8$ (say, from
$10^{11}$ to $10^3$ molecules per cm$^3$ -- to be more precise, see
Fig. \ref{fig:condensephys}).  As a consequence, it must be a good
approximation to regard $f$ to be a constant $f_0$ for calculations of
the mechanical state of the most interesting parts of a condensing
cloud core.

To justify this conclusion mathematically, define a dimensionless
measure of the time by
\be
\tau \equiv {|t|\over t_0},
\label{tau}
\ee
where $t_0$ is an arbitrary unit of time used to make the
argument of the logarithm dimensionless. To be definite, if we think of the core as having an outer boundary
that connects to a common envelope at $\varpi_{\rm ce}$, we may
choose $t_0$ to equal the time it takes a fast MHD wave traveling at speed $\Theta^{1/2} a$ to traverse the distance
$\varpi_{\rm ce}$.  In numerical terms, this would typically make $t_0 \sim 10^6$ yr.
With the definition (\ref{tau}), equation (\ref{indeq}) becomes
\be
-{\partial f\over \partial \ln \tau} +
(x + \tilde v) {\partial f \over \partial x} 
= {\epsilon \over x\tilde \sigma^2} {\partial \over \partial x} \left[  x 
\tilde \sigma \tilde b \left( {f^2 \over \sqrt{1+f^2} }\right) \right] \, .
\label{indeqtau}
\ee
We proceed now to solve the governing set of equations by an iterative process.

Begin by denoting solutions of the dynamical equations with $f$ taken
to be a fixed $f_0$ by the symbols $\tilde \sigma_0$, $\tilde v_0$,
and $\tilde b_0$.  With the substitution of equation (\ref{bdef}),
when $\tilde b = \tilde b_0$, it is then trivial to show that the
governing equation of continuity reads
\be 
\left( 1 + x {d \over d x} \right) \tilde \sigma_0(x) + 
{1 \over x} {d \over d x} 
\left[ x \tilde v_0(x) \tilde \sigma_0(x) \right] = 0 \, ,
\label{contzero}
\ee 
whereas the force equation becomes 
\be 
\left[ x + \tilde v_0(x) \right] {d \tilde v_0 \over d x} 
+ {\Theta_0 \over \tilde \sigma_0(x)} {d\tilde \sigma_0 \over d x} = 
- [1 - f_0^2] \int_0^\infty K_0 \left( {y\over x} \right)  
\tilde \sigma_0 (y) \, {y dy \over x^2} \, ,
\label{forcezero}
\ee 
with
\be
\Theta_0 \equiv {1 + 2 f_0^2 \over 1 + f_0^2} \, . 
\label{Theta0} 
\ee

Once solutions for $\tilde \sigma_0$, $\tilde v_0$, and $\tilde b_0$
have been found, we can return to equation (\ref{indeqtau}) and
replace the factor $f^2/\sqrt{1+f^2}$ in the order $\epsilon$
diffusion term by its zeroth-order approximation
$f_0^2/\sqrt{1+f_0^2}$.  We can then obtain a better estimate for $f$
by integrating the resulting {\it linear} partial differential
equation for $f$,
\be
-{\partial f\over \partial \ln \tau} +
(x + \tilde v_0) {\partial f \over \partial x} 
= \epsilon \left( {f_0^3 \over \sqrt{1+f_0^2} }\right) 
{1 \over x\tilde \sigma_0^2} {\partial \over \partial x} 
\left[  x \tilde \sigma_0 \tilde b_0 \right] \, , 
\label{diffuse}
\ee
where we have defined
\be
\tilde b_0(x) \equiv \int_0^\infty K_0\left( {y\over x}\right)
\tilde \sigma_0(y) \, {ydy\over x^2} .
\label{fieldzero}
\ee 
Note that the flux ratio $f$ = $f_0$ has been removed from the
definition of ${\tilde b}_0$ and is now included in the leading
coefficient.  In principle, one could continue to iterate solutions of
the flux-to-mass distribution from the induction equation with
solutions of the surface density (from which we can get the magnetic
field from the flux-to-mass distribution) and velocity field from the
equations governing the mass and momentum flow of the fluid to obtain
increasingly accurate numerical answers to the overall problem.  In
practice, we shall stop at the perturbative step (\ref{diffuse}).

\subsection{Homology and Self-similarity}

By introducing new scaled variables, we can transform the governing
ordinary integro-differential equations for the zeroth-order dynamics
into a universal form that is nominally independent of the numerical
value of $\lambda_0 = f_0^{-1}$.  Specifically, we adopt a scaling
transformation of the form
\be
\xi \equiv x/\sqrt{\Theta_0} ,
\label{xi}
\ee
\be
 v(\xi) \equiv \tilde v_0(x)/\sqrt {\Theta_0},
\label{hatv}
\ee 
\be
 \sigma(\xi) \equiv \tilde \sigma_0(x)(1-f_0^2)/\sqrt{\Theta_0} , 
\label{hatalpha}
\ee
where the scaling coefficients are independent of $x$. Note that the
flux ratio must obey the constraint $f_0 < 1$ (and hence $\lambda > 1$) 
for the surface densities $\sigma$ and $\tilde \sigma_0$ to be positive.  
The scaled forms of the equation of motion then become 
\be
\discrim {d v\over d\xi} = (\xi +  v) 
\left(  F + {1 \over \xi} \right) \, , 
\label{dervelgen} 
\ee 
and 
\be
{\discrim \over  \sigma} {d \sigma \over d\xi} = 
- { (\xi +  v)^2 \over \xi} -  F \, , 
\label{deralphagen} 
\ee
where the normalized force $ F$ is defined by 
\be
 F(\xi) = - {1 \over \xi^2} \int_0^\infty K_0 (\eta/\xi) 
 \sigma (\eta)  \, \eta d\eta \, . 
\label{normforce} 
\ee 
In the equations above, the discriminant $\discrim$ is given by 
\be 
\discrim \equiv (\xi +  v)^2 - 1 \, . 
\label{defdiscrim}
\ee

It is easy to see that the normalized equations (\ref{dervelgen}) and
(\ref{deralphagen}) are exactly what would have resulted if we had
looked at the outset for self-similar solutions of the un-magnetized
problem, $f_0 = 0$, $\Theta_0 =1$.  This {\it mathematical homology}
explains the decades of confusion and controversy as to what
constitutes the proper ``initial conditions'' for the latter type of
calculations (e.g., Larson 1969, Penston 1969, Shu 1977, Whitworth \&
Summers 1985, Foster \& Chevalier 1993, Andr\'e et al.~2000).  Our
group (e.g., Shu, Adams, \& Lizano 1987, Lizano \& Shu 1989, Li \& Shu
1996) has long maintained that the pivotal {\it instant} $t = 0$
represents, not an ``initial condition" where singular conditions are
reached at the origin, but rather a transitional {\it instant}
between an extended period, $t < 0$, of magnetic evolution through
flux loss via ambipolar diffusion, and another period, $t > 0$, of
dynamical collapse, infall, and star plus centrifugal disk formation.
After some initial debate, the group led by Mouschovias has come to
the same conclusion (see, e.g., Tassis \& Mouschovias 2005).  This
paper then provides the mathematical justification for the latter
point of view, and it supplies the means to select from the wealth of
``extended-contraction/runaway-condensation'' solutions for $t < 0$
advocated first by Hunter (1977) as worthy alternatives to Shu's
(1977) choice to start at $t=0$ with singular isothermal systems at
rest, after what Shu argued would be a period of subsonic evolution to 
reach such a state.  The corresponding static starting state here
reads: $ v= 0$, $ \sigma = 1/\xi$, and $ F = -1/\xi$, which provide
exact solutions of the equations (\ref{dervelgen}), (\ref{deralphagen}), 
and (\ref{normforce}), but not exactly those that we want here. 

The critical point of the flow occurs where $\discrim = 0$. In order
for the flow to pass smoothly through the critical point $\xi_\ast$,
the right-hand sides of both equations (\ref{dervelgen}) and
(\ref{deralphagen}) must vanish where $\discrim = 0$.  This
requirement defines two conditions, which act to fix
the values of $ v$ and $ \sigma$ at the critical point $\xi =
\xi_\ast$.  Using L'H$\hat{\rm o}$pital's rule, we can integrate
inwards and outwards from $\xi_\ast$.  We require that the inward
integration satisfies the inner boundary condition $ v= 0$ at $\xi =
0$. Note that only one value of $\xi_\ast$ can satisfy this
constraint. With the critical point thus specified, the outward
integration from the same point $\xi = \xi_\ast$ produces the
asymptotic behavior $ \sigma \rightarrow A \xi^{-1}$ and $ v \rightarrow
{v}_\infty$ as $\xi \rightarrow \infty$. 

\subsection{The Flux to Mass Distribution and Intermediate Asymptotics}

In reduced and scaled variables, the equation (\ref{diffuse}) can be
written
\be
-{\partial f\over \partial \ln \tau}+[\xi+v(\xi)]
{\partial f\over \partial \xi} = -
\left[{\hat \epsilon \over \xi \sigma^2(\xi)}\right] 
{d\over d\xi}\left[ \xi \sigma(\xi)F(\xi)\right] ,
\label{diffusexi}
\ee
where we have defined
\be
\hat \epsilon \equiv {\epsilon f_0^3\over\sqrt{1+2f_0^2}}.
\label{hatepsilon}
\ee
Note that the dependence of the overall scaled problem on the
parameters $\epsilon$ and $f_0$ enters explicitly only in equation
(\ref{diffusexi}) through $\hat \epsilon$.  In what follows, we shall
see that the proper formulation of an initial-value problem for the
solution of equation (\ref{diffusexi}) will connect $\epsilon$ and
$f_0$, i.e., that the flux-to-mass ratio of the central-most regions
of a condensing cloud core depends on the dimensionless rate of
ambipolar diffusion as measured by the parameter $\epsilon$.

The linear partial differential equation (\ref{diffusexi}) can be
attacked by the method of characteristics:
\be
{df\over d\xi} = \hat \epsilon {\cal N}(\xi) 
\qquad \;\; {\rm on\; the\; trajectory} \;\;  \qquad 
{d\over d\xi}(-\ln \tau) = {1\over \xi+v(\xi)} \, ,
\label{characteristics}
\ee
where
\be
{\cal N}(\xi) \equiv -{1 \over \xi \sigma^2 (\xi+v) } 
{d \over d \xi} \left[\xi \sigma F(\xi)\right]
\label{calN}
\ee
is regarded as a known function of $\xi$.   
Thus, we may define the formal integral,
\be
N(\xi) \equiv \int_1^\xi {\cal N}(\xi) \, d\xi 
\qquad {\rm and} \qquad T(\xi) \equiv \int_1^\xi {d\xi \over \xi+v(\xi)},
\label{NandT}
\ee
and we write the solution to equation (\ref{diffusexi}) as
\be
f(\xi,\tau) = f(1,\tau_1) + \hat \epsilon N(\xi) ,
\label{gensolnf}
\ee
where $\xi$ is the present position at the present time $\tau$
connected to a past (or future) position $1$ at the time $\tau_1$ on a
Lagrangian trajectory that reads in similarity coordinates:
\be
T(\xi)+\ln \tau = \ln \tau_1.
\label{tau1}
\ee
With equation (\ref{tau1}) giving $\tau_1$ as a function of $\xi$ and
$\tau$, equation (\ref{gensolnf}) yields the general solution for the
advection-diffusion equation (\ref{diffusexi}) where the term $-\hat
\epsilon N(\xi)$ gives the effect of the ambipolar diffusion relative to
a comoving observer and
$f(1,\tau_1)$ gives the effect of advection if we follow the fluid
motion assuming field freezing.

To illustrate the behavior of $f(1,\tau_1)$, we first note that the
position $\xi = \varpi/\Theta^{1/2}at_0\tau = 1$ lies just outside the
origin $\varpi = 0$ as $\tau \rightarrow 0$, whereas the same $\xi =
1$ position lies at a great radial distance $\varpi$ in the limit 
$\tau \rightarrow \infty$.  Thus, we have the generic behavior: 
\be
f(1,0) = f_0, \qquad f(1,\infty) = 1,
\ee
if the supercritical core connects to a marginally critical common
envelope.  In \S5.3, for reasonable core models, we shall find that
$N(\infty)$ is small compared to unity. [See Fig. 4 and note
$N(\infty) = N_0(\infty) -N_0(1)$.]  On the other hand, note that if
$v(\xi)$ were zero, $T(\xi)$ would equal $\ln \xi$, and the
characteristic trajectory would simply follow a line defined by $\ln
(\tau \xi) = {\rm const}$, i.e., a line of constant $\tau\xi
=\varpi/\Theta_0 at_0$ or constant $\varpi$ (because fluid elements
are not moving if $v$ were zero).  Although the inflow velocity,
$v(\xi)$, is not zero in our problem, it becomes a constant at large
$\xi$, where the term $\xi+ v(\xi)$ is dominated by $\xi$. The
relationship between $\tau_1$ and $\xi$ and $\tau$ is then given by
\be
\tau_1 \approx \xi \tau .
\ee
Thus, in the limit $\tau \rightarrow 0$ with $\xi \gg 1$, when the
moment of gravomagneto catastrophe is approached, the flux to mass
distribution as given by equation (\ref{gensolnf}) has the
approximation $f(\xi,\tau)\approx f(1,\xi\tau)$ and assumes all
intermediate values between $f(1,0) = f_0$ at small $\xi\tau \propto
\varpi$ to $f(1,\infty) = 1$ at large $\xi\tau \propto \varpi$.  In
other words, during the runaway phase of core condensation, the
function $f(\xi,\tau)$ ``freezes'' with a profile that is a function
only of the Lagrangian coordinate (which could be taken to be the
enclosed cynlindrical mass) varying monotonically from $f_0$ at the
core center to unity at the outer core boundary.

Consider now a position $\xi \ll 1$ at small but finite $\tau > 0$
after runaway condensation is in progress (which occurs roughly at
$\tau \sim 1$; see \S 5.3), with the spacetime point $(\xi,\tau)$
being connected to an initial pair $(1,\tau_1)$ near the outer core
boundary, i.e., where $f(\xi, \tau) \approx f_0$ and $f(1,\tau_1)
\approx 1$.  Equation (\ref{gensolnf}) then requires
\be
-\hat\epsilon N(0) = 1-f_0.
\label{hateps}
\ee
Together with the definition (\ref{hatepsilon}), equation
(\ref{hateps}) provides us with an eigenvalue relationship between
$f_0$ and $\epsilon$.  In other words, for given $\epsilon$, runaway
condensation occurs when ambipolar diffusion has produced a central
flux to mass rato that satisfies
\be
{(1-f_0)\sqrt{1+2f_0^2}\over f_0^3N_0(1)} = \epsilon,
\label{eigenvalue}
\ee 
where $N_0(1) = -N(0) = \int_0^1 {\cal N}(\xi)\, d\xi$ (see Fig. 4 in
\S 5.3).

Unfortunately, these properties of the general solution depend on the
seemingly innocuous assumption that ${\cal N}(\xi)$ is integrable at
$\xi$ = $0$ and $\infty$. However, as shown in the following sections,
at small $\xi$ the functions $\sigma(\xi)$, $v(\xi)$, and $F(\xi)$
approach the forms
\be
\sigma(\xi) \rightarrow \sigma_0, \quad v(\xi) \rightarrow -\xi/2, 
\qquad F(\xi) \rightarrow F^\prime(0)\xi, \qquad {\rm as} 
\qquad \xi\rightarrow 0.
\label{smalleta}
\ee
Thus, for small $\xi$, ${\cal N}(\xi)$ behaves as
\be
{\cal N}(\xi) = -{4F^\prime(0)\over \sigma_0\xi} 
\qquad {\rm for} \qquad \xi \ll 1.
\ee
In these circumstances, $N(\xi,\tau)$ will diverge as
$[4F^\prime(0)/\sigma_0]\ln \xi$ as $\xi \rightarrow 0$. The
divergence arises because in the derivation for the average value of
${\cal C}^{-1}$ in equation (B3), we have set $\partial
B_\varpi/\partial z$ equal to $(B_\varpi^+/z_0){\rm sech}^2(z/z_0)$.
This approximation is valid away from the origin, but at the origin,
$\partial B_\varpi/\partial z$ is doubly small, because not only
$B_\varpi \propto B_\varpi^+$ itself is small, but the magnetic field
is vertical near the origin so $\partial /\partial z$ is also small.
The replacement of $\partial B_\varpi/\partial z$ by something
proportional to $B_\varpi^+/z_0 \propto \sigma(\xi)F(\xi)$ accounts
for the first effect, but not the second.

As a related point, the current density $\propto (\partial B_\varpi
/\partial z-\partial B_z/\partial \varpi)$ is dominated at the origin,
not by $\partial B_\varpi/\partial z$, but by $\partial B_z/\partial
\varpi$, implying that the Lorentz force there comes mostly from the
gradient of the ``magnetic pressure'' $-\partial (B_z^2/8\pi)/\partial
\varpi$ rather than from the ``magnetic tension'' $(B_z/4\pi)\partial
B_\varpi/\partial z$.  This dominance is evident in that it is the
pressure (gas plus magnetic) that decelerates the inflow to rest at
the origin, not the tension (see eq. \ref{deceleration-origin}).  The
transition in roles of the tension versus the pressure when one moves
from the disk of the core to its central regions has implications for
the ambipolar diffusion that occurs near the origin, which the
unadulterated diffusion term $\cal N$ in equation (\ref{calN}) does
not treat correctly. Indeed, near the origin, the diffusion term
involves the second derivative $\partial^2 B_z/\partial \varpi^2$ that
translates into a term proportional to $\partial^2(\sigma f)/\partial
\xi^2$.  In a rigorous discussion, we would examine the central
regions anew and provide a proper matching of the solutions there with
those applicable for the more highly flattened regions of the cloud
core developed here. The extra radial derivative that appears via
$\partial^2(\sigma f)/\partial \xi^2$, multiplied by a small
coefficient $\epsilon$, makes possible asymptotic matching of the
inner solution to an outer or intermediate solution. In particular, it
would be possible to invoke an extra boundary condition, say,
$\partial f/\partial \xi = 0$, at the origin to ensure that $f$ is
well-behaved at the origin.  A formal attack along these lines would
involve singular perturbation theory, coupled with the introduction of
multiple length and/or time scales (Bender \& Orzag 1978).

In other words, our problem is really one of {\it intermediate
asymptotics} (Barenblatt 1996). A proper treatment would
asymptotically match the intermediate core on both an outer scale to a
common envelope, where the assumption of gravitational contraction
breaks down.  It would do the same on an inner scale to the central
region, where the assumption of a flattened configuration is invalid.
For the sake of physical clarity, we forego such a formal study in
this paper, and join the intermediate core to a common envelope only in
terms of its magnetic connection and not in its the mechanical
considerations.  We also incorporate the region near the origin into
the intermediate analysis by fixing the problem presented above, as
well as one that will appear later, with simple procedures that are
physically rather than mathematically motivated.  We defer to \S 5.3
therefore our prescription for making ${\cal N}(\xi)$ regular at the
origin.

\section{SHEET CONDENSATION SOLUTION}

\subsection{Gravity Formulation in Terms of the Enclosed Mass}

In order to obtain an approximation to the condensation solution for
$t < 0$, we can make the assumption that the potential is given by the
monopole associated with the enclosed mass.  Although it is possible
to develop this approximation as the first term of a general multipole
expansion (see Appendix E), we prefer a motivation based on physical
intuition.  We start by defining the cylindrically enclosed mass $M
(\varpi, t)$ in dimensional units:
\be
M (\varpi, t) = \int_0^\varpi 2 \pi r dr \Sigma(r,t) \, . 
\label{interiormass}
\ee
The statement that the enclosed mass is conserved if we follow the
motion of mass annuli,
\be
{\partial M\over \partial t}+u{\partial M\over \partial \varpi} = 0 \, ,
\label{massconsv}
\ee
may be regarded as an integrated form of the continuity equation
(\ref{continuityeq}).  If we use the dimensionless variables defined
by equations (\ref{dimensionless}), then equations (\ref{interiormass}) 
and (\ref{massconsv}) become 
\be
M(\varpi,t) = {a^3 \over G} |t| \tilde m(x) \qquad {\rm where} 
\qquad \tilde m(x) = \int_0^x \tilde \sigma \, xdx \, , 
\ee
\be
-\tilde m+(x+\tilde v){d\tilde m\over dx} = 0,
\label{tildemass}
\ee
where we have assumed the case where $t < 0$.  We employ the same
scaling transformation as before (see eqs. [\ref{xi}] --
[\ref{hatalpha}]) to express the equations in terms of the variables
$\xi$, $ v$, and $ \sigma$.  The dimensionless enclosed mass now
becomes
\be
\tilde m(x) = {\Theta_0^{3/2}\over 1-f_0^2}\, m(\xi) \, , \qquad 
{\rm where} \qquad m(\xi) \equiv \int_0^\xi \sigma \, \xi d\xi.
\label{interiormassx}
\ee
The differential version of the last equation, 
\be
{dm\over d\xi} = \xi \sigma,
\label{masselement}
\ee
may be combined with the scaled version of equation (\ref{tildemass}),
\be
m = (\xi+v){dm\over d\xi} = \xi(\xi+v)\sigma,
\label{massconsvxi}
\ee
to express the monopole approximation for the force in the form 
\be
F(\xi) = -{m\over \xi^2} = -{1\over \xi} (\xi+v)\sigma .
\label{monopoleforce}
\ee

As a check, notice that $\xi$ times the equation of continuity,
written in the usual fashion as the scaled version of equation
(\ref{contzero}),
\be
\xi {d\over d\xi}(\xi\sigma) +{d\over d\xi}\left( \xi \sigma v\right) 
= -\xi\sigma +{d\over d\xi}[(\xi+v)\xi\sigma] = 0,
\label{xicont}
\ee
is simply the derivative of equation (\ref{massconsvxi}) with respect
to $\xi$.  With the force $F$ reduced to a local expression, equations
(\ref{dervelgen}) and (\ref{deralphagen}) become the following coupled 
set of first-order, nonlinear, ordinary differential equations:
\be
\discrim {d { v} \over d\xi} = {(\xi + { v}) \over \xi} 
\left[ 1 -  { \sigma} (\xi + { v}) \right] \, , 
\label{dervel} 
\ee 
\be
\discrim {d { \sigma} \over d\xi} = (\xi + { v}) 
{{ \sigma} \over \xi} \left[ { \sigma} - 
(\xi + { v}) \right] \, , 
\label{deralpha} 
\ee
where the discriminant $\discrim$ is given by equation
(\ref{defdiscrim}).  

\subsection{Critical Points} 

At the critical points $\xi_\ast$, $\discrim$ = 0, so the right hand
sides of equations (\ref{deralpha}) and (\ref{dervel}) must also
vanish.  This condition thus determines the value of the density field
at the critical point,
\be
{ \sigma} (\xi_\ast) = 1 \, . 
\ee 
If we expand around the critical point, the leading order corrections 
have the forms 
\be 
\xi = \xi_\ast + \delx \, , \qquad 
{ v} = v_\ast + { v}_1 \delx \, , \qquad {\rm and} \qquad 
{ \sigma} =  \sigma_\ast + { \sigma}_1 \delx \, . 
\label{critexpand}
\ee 
Using these expressions in the equations of motion and keeping only the 
leading order terms, we can find the field derivatives at the 
critical point:
\be
{ v}_1 = - {1 \over 2} \pm {1 \over 2 \xi_\ast} 
\left[ (\xi_\ast - 1)^2 + 1 \right]^{1/2} 
\qquad {\rm and} \qquad  
{ \sigma}_1 = {1 \over 2 \xi_\ast} \left\{ \xi_\ast - 2 
\mp \left[ (\xi_\ast - 1)^2 + 1 \right]^{1/2} \right\} \, . 
\ee 
Appendix D generalizes this procedure for arbitrary forms of $F(\xi)$.

\subsection{Limiting Forms} 

In the limit $\xi \to \infty$, the force equation allows for an
asymptotic solution for the surface density, namely
\be 
{\sigma} = {A \over \xi} \, , 
\label{outeralf} 
\ee 
where $A$ is a constant.  With this form for the surface density, the
asymptotic behavior for the velocity can be found,
\be
{v} (\xi) \to {v}_\infty + {A - 1 \over \xi} \, , 
\label{outervel} 
\ee 
where ${v}_\infty$ is a constant. The second correction term goes to
zero as $\xi \to \infty$, but the falloff is slow.  As a result, the
inflow velocity will in general be nonzero but small even at the
pivotal instant of gravomagneto catastrophe (Allen et al. 2003,
Fatuzzo et al. 2004); this result, in turn, implies that the
subsequent infall rates will be larger than for cases where the
starting states are in exact hydrostatic equilibrium.

In the limit $\xi \to 0$, we want to enforce the boundary 
conditions 
\be 
{v} \to 0  \, \qquad {\rm and} \qquad 
{\sigma} \to {\sigma}_0 = {\rm constant} \, . 
\label{innerbc} 
\ee 
The solution for the velocity field near $\xi = 0$ has the dependence 
\be
{v} (\xi) = - {1 \over 2} \xi \, . 
\label{innervel} 
\ee 
With this limiting expression for $v$, the density field takes the form 
\be
{\sigma} (\xi) = 
{{\sigma}_0 \over 1 + {\sigma}_0 \xi/2} \, . 
\label{inneralf} 
\ee 
Notice that this particular function becomes ${\sigma} = 2/\xi$ 
at large values of the scaled similarity variable $\xi$, whereas the 
asymptotic limit of the density field has the form ${\sigma} = A/\xi$. 

Unfortunately, the behavior near the origin is less than perfect in a
model of starless cloud cores as flat sheets. The monopole associated
with a sheet produces a net scaled force (gravitational plus magnetic
tension) that approaches the form $F \to - \sigma_0/2 = -2.524$ as
$\xi \rightarrow 0$ (see eq.~[\ref{monopoleforce}] and below).  The
origin of this behavior rests with the unscaled reduced force behaving
as $\tilde F = -(1-f_0^2)\tilde m_0/x^2$ with $\tilde m_0 = \tilde
\sigma_0 x^2/2$ if the surface density approaches a constant $\tilde
\sigma_0$ in the limit $x \rightarrow 0$. In other words, with
constant surface density, the mass enclosed within a cylindrical
radius scales as the square of that radius, which cancels the inverse
square law of Newtonian gravity for monopoles (and Amp\'erian
magnetism for split monopoles).  Hence, unlike the classic freshman
physics problem of the gravitational field inside a sphere of uniform
volume density, the corresponding force in an axisymmetric flat sheet
of uniform surface density does not go to zero, even when we approach
the origin!

However, by symmetry considerations, the radial force must vanish
right at the origin for any axisymmetric mass and current distribution
that is regular there.  Appendix E shows that a jump in the physical
behavior as one changes from being at the origin to being slightly off
it is generic to {\it any} order in a general multipole expansion of
an axisymmetric sheet.  On the other hand, incompletely flattened,
magnetized, isothermal, disks/toroids have scaled reduced half-height
$\zeta_0$ and dimensionless {\it volume} density of the form $\alpha =
(\sigma/2\zeta_0) {\rm sech}^2 (\zeta/\zeta_0)$ (see Appendix A).  If
the volume density is regular at $\xi = 0$ (rather than diverging as
$\xi^{-2}$, which applies only at the pivotal instant $t = 0$), then
$F \propto \xi \rightarrow 0$ as $\xi \rightarrow 0$ (see Appendix F).
Accounting for the finite thickness of actual molecular-cloud cores
thus cures the unphysical situation at the origin. As a result, a
solution that enforces physical boundary conditions at the origin
based on the fluid's reaction to a {\it sheet} monopole is clearly
blemished.  In the interest of obtaining practical and useful results,
however, we defer further discussion of this imperfection until \S 5.

\subsection{Sheet Monopole Solution} 

\begin{figure}
\figurenum{1}
{\centerline{\epsscale{0.90} \plotone{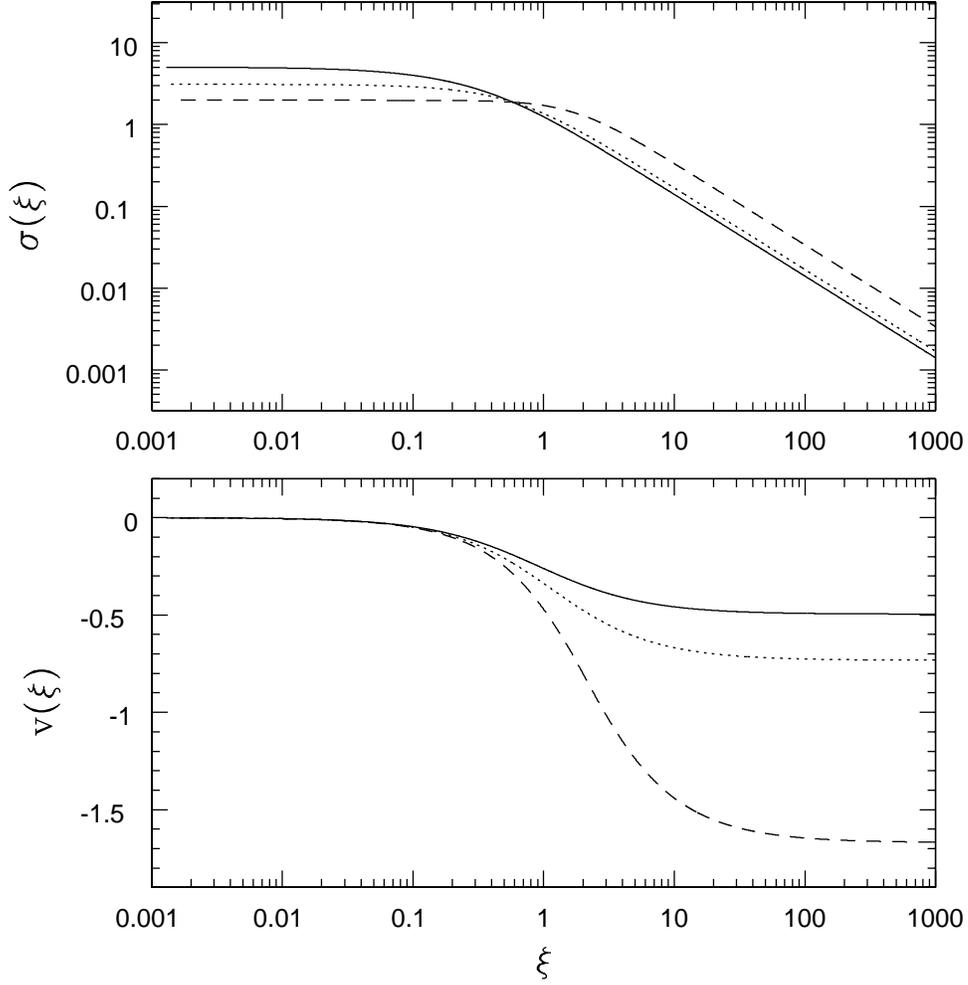} }} 
\figcaption{ Reduced fluid fields for the pure monople, monopole plus
quadrupole, and full sheet solutions plotted, respectively, as the
solid, dotted, and dashed curves.  The solutions for reduced and
scaled surface density $\sigma(\xi)$ and velocity $v(\xi)$ are given,
respectively, by the upper and lower panels, with the limiting values
$\sigma \rightarrow \sigma_0$ and $v\rightarrow 0$ as $\xi \rightarrow
0$; $\sigma \rightarrow A/\xi$ and $v \rightarrow v_\infty$ as $\xi
\rightarrow \infty$. The parameter values are tabulated in Table 1. } 
\label{fig:monopole} 
\end{figure}

With the preliminaries in place, we can now find the critical points,
and integrate both inward toward $\xi = 0$ and outward to large $\xi$
to find the solutions for the reduced and scaled density and velocity
fields (in this monopole approximation). In the usual case, these two
coupled ODEs would require two boundary conditions to specify a
solution. In this setting, equations (\ref{innerbc}) supply the inner
boundary conditions on $ v$ and $ \sigma$, but we do not know the
correct value ${\sigma}_0$ to enforce. However, this problem contains
an additional constraint, namely, that the flow must pass smoothly
through the critical point $\xi_\ast$ (specifically, the fluid fields
$ v$ and $ \sigma$ must be continuous at $\xi_\ast$, but their
derivatives need not be). Since each starting value of ${\sigma}_0$
would lead to a different value of $ \sigma$ at the critical point,
only one value ${\sigma}_0$ allows for smooth flow. To find this
value, we start the integration at a possible critical point and
integrate inwards toward $\xi$ = 0. By requiring that the solution
satisfy the inner boundary condition on $ v$, we can iterate the
starting point until we find the correct value of the critical
point. With this value specified, we then integrate outward from the
critical point. With no further quantities to specify, this
integration thus determines both $A$ and ${ v}_\infty$.  The resulting
solution is shown in Figure \ref{fig:monopole} (along with solutions 
from the following section).

The solutions for $v$ and $ \sigma$ follow the limiting forms found
analytically in \S 4.3. From the numerical solution, we can find the
values of the parameters appearing in the analytic forms: $\xi_\ast =
1.294$, ${\sigma}_0 = 5.03$, $A = 1.40$, and $-v_\infty = 0.495$.  We
can combine the inner and outer limiting forms for the density profile
to construct an approximate solution,
\be
\sigma_A (\xi) = 
\left({\sigma_0 \over 1 + \sigma_0 \xi/2}\right) 
\left( {1 + A \xi/2 \over 1 + \xi}\right) , 
\label{alphaprox} 
\ee 
that agrees well with the numerical solution with an RMS error over
the range of $\xi$ shown in Figure \ref{fig:monopole} of only
$\sim1\%$. Similarly, we can fit the velocity field with the form 
\be
v_A (\xi) = {v_\infty \xi \over \xi + 2 |v_\infty| } \, , 
\label{vapprox}
\ee
where this result agrees with the numerical solution with an RMS 
error of $\sim2$\%. 

One of the most important quantities resulting from this calculation
is the value of the nonzero velocities in pre-collapse cores, i.e.,
cores that have not yet produced protostars at their centers. The
physical value of this inward speed is given by $u_\infty = a v_0 (\xi
\to \infty) = a {v}_\infty \sqrt{\Theta_0}$. Since the flux to mass
ratio $f_0$ must lie in the range $0 \le f_0 \le 1$, the correction
parameter $\Theta_0$ is confined to the range $1 \le \Theta_0 \le
3/2$.  As a result, the head-start velocity is constrained to lie in
the range
\be
0.495 \le - {u_\infty \over a} \le 0.606 \, , 
\ee
where $a$ is the isothermal sound speed (and the minus sign denotes
inward velocities).  These values are consistent with, although
perhaps slightly smaller after projection, than the extended infall
velocities observed in starless molecular-cloud cores (e.g., Lee et
al. 2001, Harvey et al. 2002).

Another important physical quantity in this problem is the size of the
region of nearly constant density in the core center.  The similarity
solution (Fig. \ref{fig:monopole}) shows that the surface density 
$\sigma$ has zero slope in the center and steepens to the form $\sigma
= A/\xi$ in the outer regime. We can thus define the outer boundary of
the core region to be the location where the index $p \equiv -(\xi/\sigma) 
(d \sigma/d\xi)$ = 1/2. Using the approximate form given by equation
(\ref{alphaprox}), we find that the outer boundary of the core region
occurs at $\xi_{1/2} \approx 0.313$.  The physical location of this
boundary is given by
\be
\varpi_{1/2} \approx \, 0.313 \, a \, |t| \, \sqrt{\Theta_0} \, , 
\ee 
where the column density falls to 0.519 of its central value. 

To summarize, before $t = 0$, condensing, magnetized molecular cloud
cores display a finite region of nearly constant central surface
density that makes them mimic static Bonnor-Ebert spheres.  In
actuality, however, the surrounding regions of the core are in a state
of extended contraction at a significant fraction of the isothermal
sound speed $a$.  In the limit $t\rightarrow 0^-$, this central region
loses its finite extent and the core attains a pure-power law
configuration, $\Sigma \propto \varpi^{-1}$. that corresponds ideally
in three dimensions to a flattened singular isothermal toroid, $\rho
\propto r^{-2}R(\theta)$ in spherical polar coordinates. The core
subsequently goes into true dynamical collapse onto a central
protostar.

\section{BEYOND THE SHEET MONOPOLE APPROXIMATION} 

Before we consider the collapse solution for $t > 0$ (see \S 6), we
consider the three shortcomings in our treatment of the gravity of a
flattened core by the sheet monopole approximation.

\noindent
[1] For a flattened core, the monopole approximation represents only
the first term in a more general multipole expansion (see Appendix E).
These higher-multipole terms convey two types of corrections: Inner
multipoles correct for the fact that matter in a flat sheet interior
to the field point $\xi$ is (on average) closer to $\xi$ than if the
enclosed mass were placed at the core center.  Outer multipoles
correct for the gravitational pull of the matter in the sheet outside
of the field point $\xi$.  The physical description for the action of
currents is more complicated because we have used a scalar potential
rather than the vector potential to describe the magnetic field (see
Li \& Shu 1997), but the consequence for the magnetic tension is the 
same except the tension force acts in opposition to self-gravity. 

\noindent
[2] The aspect ratio, which is given by 
\be
{z_0\over \varpi} = {2\over (1+f_0^2) x \tilde \sigma} = 
\left[ {2(1-f_0^2)\over 1+2f_0^2}\right]\left({1\over \xi \sigma}\right) \, ,
\label{aspectratio}
\ee
cannot be small compared to unity in the limit $\xi \rightarrow 0$.
Indeed, even at large $\xi$, $z_0/\varpi = 2(1-f_0^2)/A(1+2f_0^2)$,
which equals $1.4 (1-f_0^2)/(1+2f_0^2)$ for the sheet monopole
solution, and is not small unless $f_0$ is close to unity.  The
assumption that the cloud core is highly flattened is egregiously
violated in the central regions, precisely where the surface density
profile flattens instead of continuing inwards as $\sigma = A/\xi$.
For the monopole solution, at $\xi = \xi_{1/2}$ we have $z_0/\varpi =
2.44(1-f_0^2)/(1+2f_0^2)$, which has a value 1.22 for a typical $f_0 = 1/2$. 

\noindent
[3] The regions at large $\xi$ are not fully isopedic with constant $f
= f_0$ (see \S 5.3). If $f$ is an increasing function of $\xi$, then
the force of magnetic tension becomes increasingly strong relative to
the force of self-gravity, instead of maintaining a constant ratio
(with opposite signs), as is true in the inner core.  Growing magnetic
support against self-gravitation as the envelope is approached will
presumably also reduce the induced inflow velocities.

We now discuss how these shortcomings may affect a peculiar aspect of
the sheet monopole solution obtained in \S 4.  We found that the flow
properties are completely defined by the behavior of the gas near the
origin $\xi = 0$.  In particular, Appendix C proves that the velocity
profile is monotonic, which implies that if inflow occurs at any point
in the self-similar system, then (for $t < 0$) the flow must pass
through a critical point where $v = 1-\xi$, and approach the form $v =
-\xi/2$ near the origin.  The latter behavior during the cloud-core
condensation stage has nothing to do with gravitational minus tension
forces.  It represents the deceleration of an initially inwardly
directed velocity by the pressure forces.  Independent of how the
force $F$ behaves, as long as it does not diverge at the origin,
equation (\ref{dervelgen}) implies at small $\xi$,
\be 
[-1]{dv\over dx} = {1\over \xi}(\xi+v), 
\label{deceleration-origin} 
\ee
where the $-1$ remaining in the discriminant comes from the pressure
gradient.  The above equation has the solution $v = -\xi/2$ if the
velocity $v$ vanishes at the origin.

In itself, the above result is not particularly ominous.  But smooth
passage through the critical point also determines the solution that
is reached at asymptotic infinity, in particular, the values of the
head-start velocity $v_\infty$ and the surface-density coefficient
$A$.  How is it possible for the conditions near the center of a
condensing cloud core to dictate how the core connects at asymptotic
infinity to the cloud envelope?  Isn't that putting the cart before
the horse?  In principle, if there is enough time and the inflow is
sub-magnetosonic, as is the case with the monopole solution, then one
could imagine the central regions to have a magnetohydrodynamic
influence on the outer regions.  However, we are not guaranteed such
simple behavior in every circumstance, and naive treatments of the
core gravity can produce super-magnetosonic condensation speeds (see
below).  Wouldn't such solutions be unstable to shock formation as the
pressure forces attempt to bring the inflow to a halt at the core
center?  (Compare this question with previous criticism [Shu 1977] of
the Larson-Penston solution, which is over-dense and supersonic by
even larger margins, and the placement of the Larson-Penston solution
in the context of {\it champagne} outflows [Shu et al. 2002].)

Slow core condensation by ambipolar diffusion avoids the above
paradox.  As flux is lost from the central regions to the outer
regions above (to reach typical values of $f_0 \approx 1/2$), the
inner core begins a stage of extended contraction at a fraction of the
magnetosonic speed -- as is seen in both numerical simulations (e.g.,
Basu \& Mouschovias 1994) and observations (e.g., Lee et al. 2001,
Harvey et al. 2002).  If the leakage of the flux is slow, as it must
be because the envelope of a typical molecular cloud is too
well-ionized to allow rapid motion of neutrals past ions, then the
condensing core never loses so much support that its surface-density
coefficient in the outer parts becomes substantially greater than the
equilibrium value $A = 1$.  Without large over-densities $A-1$, it is
not possible to generate large head-start velocities $v_\infty$ --
unless one has an over-idealized force calculation for $F(\xi)$ in the
central regions by assuming the region is highly flattened when it is
not.

With the above comments in mind, we extend the flattened monopole
approximation by two different methods.  In the first method, we
retain the sheet approximation but compute the force in its full form
(see Appendix E):
\be
F(\xi) = \int_0^{\infty}K_0\left({\eta\over \xi}\right)
\left[ {A\over \eta} -\sigma(\eta)\right]\, \eta d\eta -{A\over \xi}.
\label{fullforce}
\ee
which is mathematically identical to equation (\ref{normforce}).  
The difference surface density $A/\eta -\sigma(\eta)$ is everywhere
positive but rapidly goes to zero much outside of the central core
$\eta \gg \xi_{1/2}$.  Hence, the integral may be truncated at a
reasonable upper limit without compromising numerical accuracy.
Appendix D gives a formal description of the solution procedure for
such arbitrary forms of $F(\xi)$.

In the second method, we retain the interior monopole approximation,
but compute the force in a modified form.  We begin by defining the
reduced scaled half-thickness
\be
{z_0\over \Theta_0^{1/2}a|t|} = \left[{2(1-f_0^2)\over 1+2f_0^2}\right]
{1\over \sigma(\xi)} \equiv \zeta_0(\xi).
\label{zeta0}
\ee
We then replace $\xi$ in the denominator of equation
(\ref{monopoleforce}) by $\sqrt{\xi^2+\zeta_0^2(\xi)}$ on the
heuristic basis that the latter is a truer measure of the distance
between the field point and a typical interior source point in an
incompletely flattened cloud core.  The softened-monopole force
$F(\xi)$ now reads
\be
F(\xi) = -{\sigma (\xi+v)\over \sqrt{ \xi^2+\zeta_0^2(\xi)}}.
\label{modifiedmonopole}
\ee
Note that equation (\ref{modifiedmonopole}) produces a force law
that is proportional to $\xi$ for small $\xi$ where $v \approx -\xi/2$
and $\sigma \approx \sigma_0$.  Notice also that that the homology
with unmagnetized systems has disappeared; cases with different $f_0$
produce different reduced scaled variables. In partial compensation,
forces of the form (\ref{modifiedmonopole}) are directly integrable by
a slight modification of the method discussed in \S 4.  Finally,
notice that the implied ``enclosed mass'' no longer corresponds to the
cylindrical value because in a quasi-spherical geometry, it is more
appropriate to consider a central region with constant {\it volume}
density, which still integrates to a constant surface density in the
complete vertical direction, but does not behave at small $\xi$ as a
sheet of matter and current.

\subsection{Full-Gravity Corrections in a Sheet}

The case when $F(\xi)$ is the full gravity of a sheet is shown as the
dashed line in Figure \ref{fig:monopole}.  Note that the solution
has basically the same form as the sheet monopole solution (solid curves),
but exhibits somewhat different values of the defining constants.  The
critical point in the flow shifts outward to $\xi_\ast \approx 1.714$;
the central density $\sigma_0$ becomes somewhat smaller $1.98$, while
the asymptotic density coefficient $A$ increases to $3.43$, resulting
in the larger (supermagnetosonic) head-start speed $-v_\infty \approx
1.67$.  For reference, the intermediate case for the monopole plus
quadrupole corrections in a sheet geometry is plotted as dotted curves. 
We may attribute the full-gravity and monopole-plus-quadrupole results
to net inward gravitational fields that have increased in the outer
regions and decreased in the central regions relative to a pure
monopole.

\subsection{Softened Monopole Gravity}

Unfortunately, the approach of the previous subsection has its own
difficulty in the central regions because the approximation that the
disk is geometrically thin leads to an unphysical, non-vanishing,
force $F(\xi)$ near the origin before a protostar has formed there.
Equation (\ref{fullforce}) shows explicitly how tricky it is even to
conclude that $F(\xi)$ equals a constant at the origin rather than
diverging as $1/\xi$ when we model a cloud-core with a power-law
envelope and a non-singular center as a flat sheet.  

The behavior at asymptotic infinity for the softened monopole is
similar to the sheet monopole, but it is somewhat different near the
origin. The velocity field has the same form as before, i.e.,
\be
v \rightarrow -{1\over 2}\xi \qquad {\rm as} \qquad \xi \rightarrow 0 \, . 
\ee
The density field approaches the form
\be 
\sigma(\xi) = \sqrt{c_0} \left[ 1 - \left(1 - c_0/\sigma_0^2 \right) 
{\rm e}^{-\xi^2/4} \right]^{-1/2} \, , 
\ee
where we have defined 
\be
c_0 = (1 - f_0^2)/(1 + 2 f_0^2) \, . 
\ee 
Note that, instead of a finite value, the gas pressure gradient is now
zero at the origin, reminiscent of quasi-hydrostatic equilibrium.

\begin{figure}
\figurenum{2}
{\centerline{\epsscale{0.90} \plotone{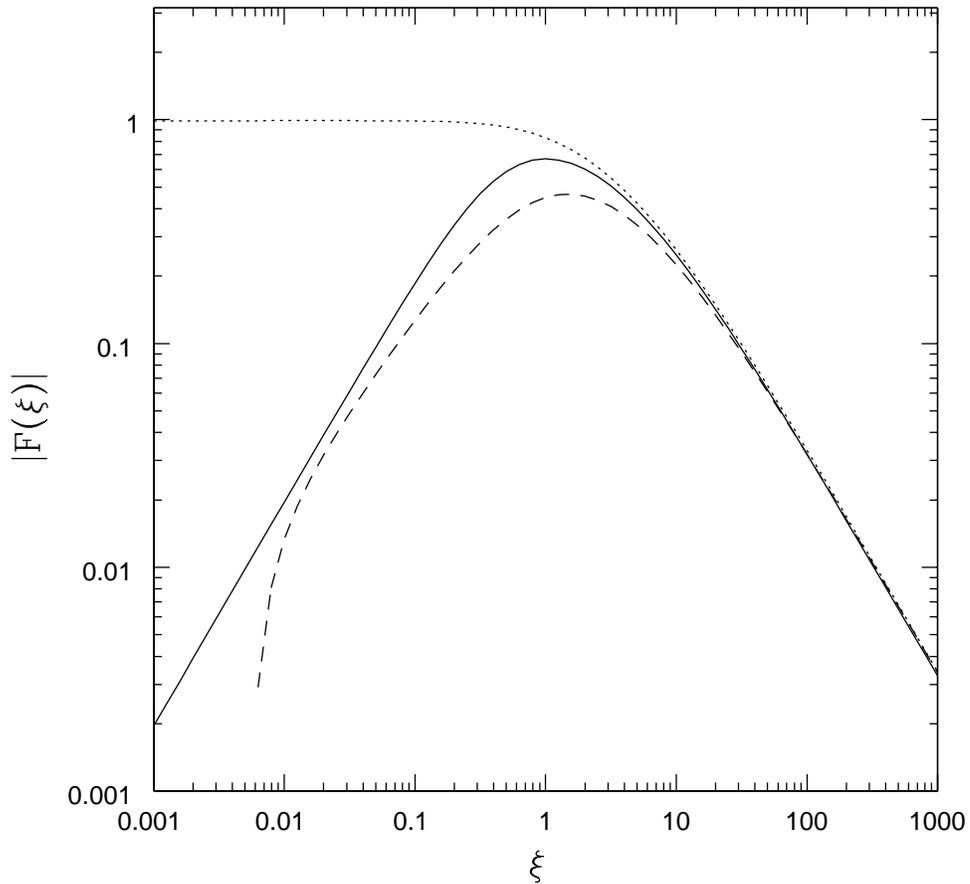} }} 
\figcaption{ Comparison of the force profiles calculated from the full
sheet gravity (dashed curve), the sheet monopole approximation (dotted
curve), and the softened monopole treatment using $f_0$ = 1/2 (solid
curve). For a fair comparison, all three force profiles are calculated
using the same surface density and velocity as obtained in \S 5.1. }
\label{fig:forcefun} 
\end{figure}

Figure \ref{fig:forcefun} shows the different approximations for the
force prescription used in this paper. The dashed curve shows the
force law for the case of full sheet gravity.  Since its force changes
sign at small $\xi$ (the net force as $\xi \rightarrow 0$ is constant
but directed {\it outward}), we cannot display the innermost region in
a log-log plot.  The dotted curve shows the force profile resulting
from the sheet monopole approximation for the same surface density and
velocity profile that resulted from the case for the full sheet
gravity. This normalization is needed to make a fair comparison of the
three techniques.  Note that the sheet monopole force approaches a
constant value (now directed {\it inward}) in the limit $\xi \to 0$,
which leads to the difficulties discussed earlier.  Finally, the solid
curve shows the force profile calculated using the same rules but with
the softened monopole approximation for the case $f_0 = 0.5$.  By
construction, its force law has a linear form at small $\xi$,
vanishing at the origin, and then joins onto the standard profile at
larger values of $\xi$.

\begin{figure}
\figurenum{3}
{\centerline{\epsscale{0.90} \plotone{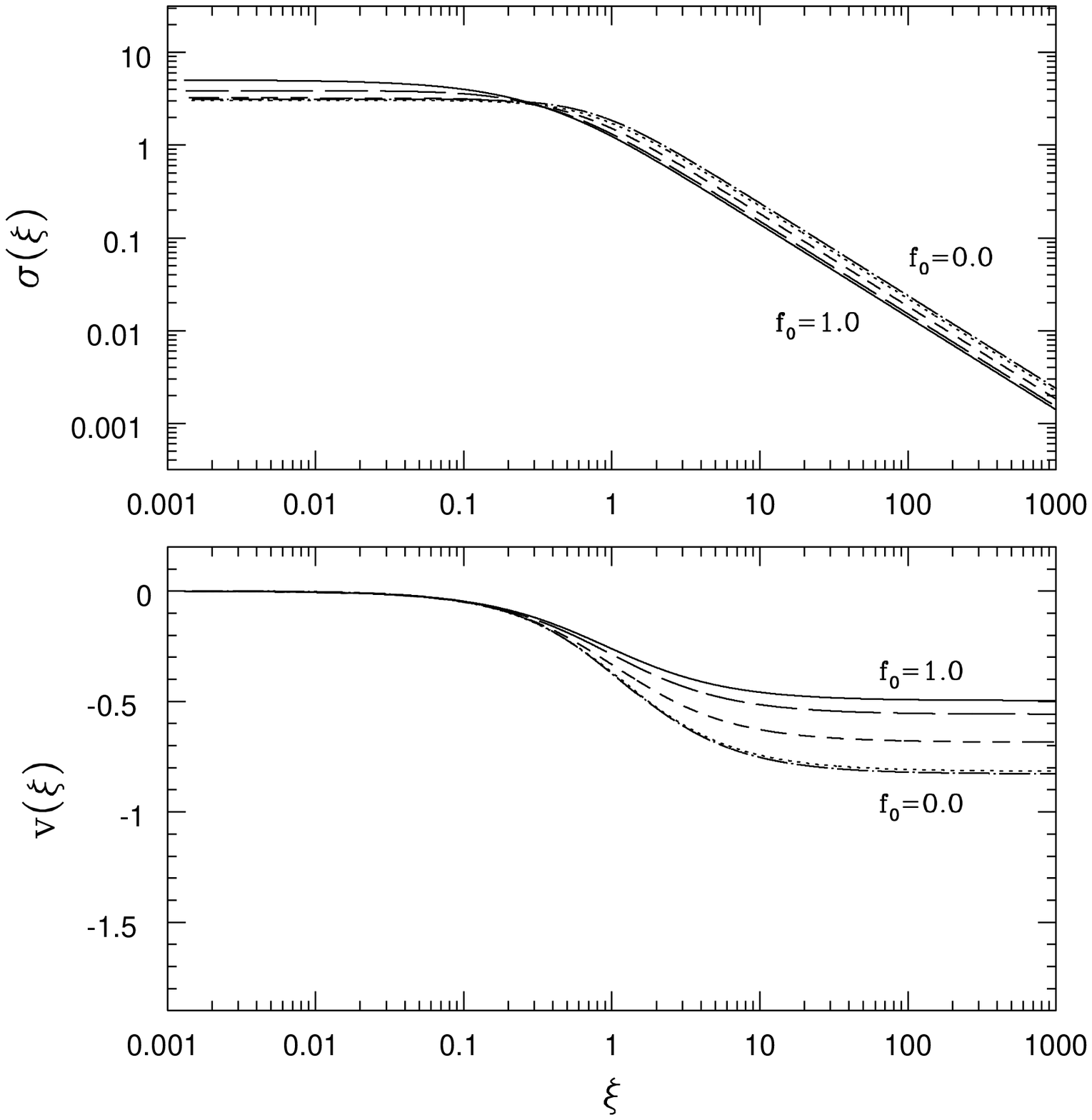} }} 
\figcaption{ Reduced fluid fields for the softened monopole solutions. 
The velocity field is shown in the lower panel and the density field is 
shown in the upper panel. In each case, the curves correspond to varying 
degrees of softening, as determined by the zeroeth order flux to mass 
ratio: $f_0$ = 1 (solid), $f_0$ = 0.75 (long dashes), $f_0$ = 0.50 (dashes), 
$f_0$ = 0.25 (dots), and $f_0$ = 0.0 (dot-dashed). } 
\label{fig:monosoft} 
\end{figure}

To study the possible range of head-start velocities, and to assess
the meaning of the artificially inflated values resulting from the
anomalies of all forms of sheet gravity at the origin, we use the
modified monopole prescription of equation (\ref{modifiedmonopole}) to
find additional condensation solutions.  The resulting solutions can
also be characterized by the parameters $\sigma_0$, $A$, and
$-v_\infty$, which are tabulated in Table 1 for varying values of
$f_0$. Figure \ref{fig:monosoft} plots the associated functions
$\sigma (\xi)$ and $v(\xi)$. Notice that as the dimensionless
flux-to-mass ratio $f_0$ decreases, the degree of softening becomes
larger, the effective scale height of the density distribution in the
radial direction becomes larger, and the critical points of the flow
move outward. The flow solutions in the outer regime are specified by
the parameters $v_\infty$ and $A$, which grow larger with decreasing
$f_0$: The softer gravity allows for more extended density profiles
and hence larger $A$, and pulls inward less strongly to allow for
greater head-start velocities $v_\infty$ (see Table 1 and Figure
\ref{fig:monosoft}).  The important point, however, is that the
head-start speed $-v_\infty$ in each case is {\it sub-magnetosonic}
(i.e., $|v_\infty| < 1$).  

Notice also that in the limit $f_0 \to 1$, the softening parameter
$\zeta \to 0$, and we recover algebraically the unsoftened monopole
approximation.  As a result, the $f_0 \to 1$ solution agrees with the
original monopole solution (see Fig. \ref{fig:monosoft} and Table 1).
This case is to be regarded as the limiting case where $f_0 \to 1$
from {\it below}.  In this limiting procedure, the weak effective
gravity is compensated by the large dimensional surface density, and
thus produces roughly the same collapse time scale (see eqns.
[\ref{coremass}] and [\ref{mdotphys}]): $M_{\rm core}/{\dot M} =
(A/m_0)(\varpi_{\rm ce}/\Theta^{1/2}a)$ for cores with different
values of $f_0$.

\begin{table}
\caption{Parameters for Diffusion Epoch Solutions} 
\medskip 
\begin{center}
\begin{tabular}{lcccc} 
\hline 
\hline 
Model & $\xi_\ast$ & $\sigma_0$ & $-v_\infty$ & A \\ 
\hline
\hline
Monopole   & 1.294 & 5.03 & 0.495 & 1.40 \\ 
Quadrupole & 1.407 & 3.12 & 0.732 & 1.68 \\ 
Full Sheet Gravity & 1.714 & 1.98 & 1.67 & 3.43 \\ 
\hline 
$f_0$ =  0.0   &  1.467 & 3.08 & 0.833 & 2.35 \\ 
$f_0$ =  0.25  &  1.455 & 3.04 & 0.815 & 2.21 \\  
$f_0$ =  0.50  &  1.393 & 3.22 & 0.684 & 1.83 \\ 
$f_0$ =  0.75  &  1.328 & 3.85 & 0.556 & 1.52 \\ 
$f_0$ =  1.00  &  1.294 & 5.03 & 0.495 & 1.40 \\ 
\hline 
\hline 
\end{tabular}
\end{center} 
\end{table}

Since a realistic treatment should include both of the effects
discussed in this subsection and the previous one, we expect the
actual results for $\sigma_0$, $A$, and $-v_\infty$ to be given by a
combination of the two types of models.  Regarding this issue, one
should remember that the first three entries in Table 1 refer to the
{\it same} sheet model; for this first model type, the full sheet
gravity case is nominally the most complete (excluding the central
regions of the core).  Models of the second type are characterized by
their $f_0$ values, which vary from $f_0$ = 0 for the unmagnetized
spherical limit to $f_0$ = 1 for the highly flattened, critically
magnetized, case.  For $f_0=1$, the disk thickness and the
``softening'' due to it disappear, and the softened monopole becomes
identical to the sheet monopole, i.e., they both lack the full
complement of higher multipoles that is present implicitly in the full
sheet gravity model.  Nevertheless, when $f_0 \neq 1$, Figure
\ref{fig:forcefun} indicates that the softened monopole model mimics
well the best characteristics of the full sheet gravity model while
adopting none of its pathologies near the origin.  For astrophysical
applications, we thus regard the results presented in the lower
portion of Table 1 to be more reliable than those of the upper
portion.  For any of these cases, the associated surface density and
velocity field may be reasonably approximated with the fitting
formulae (\ref{alphaprox}) and (\ref{vapprox}) for $\sigma_A(\xi)$ and
$v_A(\xi)$.

\subsection{Attachment of Condensing Core to Common Envelope}

We now address the solution for flux-to-mass ratio $f(\xi,\tau)$. Note
that it is possible to discuss the effects of ambipolar diffusion only
for the softened monopoles of \S 5.2 where we have made sure that
$F(\xi)$ goes linearly to zero as $\xi\rightarrow 0$. The magnetic
tension force $\propto -f_0^2F(\xi)$ acts on the ions (but not the
neutrals) and drives ambipolar diffusion via the term ${\cal N}$ in
equation (\ref{characteristics}), This term would be badly divergent
at the origin if $F(\xi)$ went to a constant there, rather than
vanishing as a linear function of $\xi$.  As it is, however, the
integral of ${\cal N}$ is still logarithmically divergent, as already
discussed in \S 3.4.  To make ${\cal N}$ even better behaved, without
a lot more analysis, which would be an onerous investment when weighed
against the limited enlightenment such an effort would yield, we adopt
the simple procedure of modifying ${\cal N}$ in the following manner.

Consider anew the derivation of Appendix B but include from the 
start the entire current:
\be 
{\partial B_z \over \partial t} + {1 \over \varpi} 
{\partial \over \partial \varpi} \left( \varpi B_z u \right) = {\cal D},
\ee
where
\be
{\cal D} \equiv {1 \over \varpi} {\partial \over \partial \varpi} \left[ 
{\varpi B_z^2 \over 4 \pi \gamma \rho \rho_i} \left(
{\partial B_\varpi \over \partial z} - 
{\partial B_z\over \partial \varpi}\right) \right] .
\ee 
We replace $\partial B_\varpi/\partial z$ by $(B_\varpi^+/z_o){\rm
sech}^2(z/z_0)$ and $\partial B_z/\partial \varpi$ by $2\pi G^{1/2}
\partial (f\Sigma)/\partial \varpi$.  Thus, where we see $B_\varpi^+$
we need to add $- z_0 f_0\partial \Sigma/\partial \varpi$ times some
coefficient that represents a thickness correction factor.  In net,
the modified form for the diffusion source term can now be written as
\be
{\cal N}_{\rm L} = -{1\over \xi \sigma^2 (\xi+v)}
{d\over d\xi}\left\{\xi \sigma \left[ {\cal M}F+
{\cal S}\left({1-f_0^2\over 1+f_0^2}\right) 
{1\over \sigma}{d\sigma\over d\xi}\right]\right\},
\ee
where $\cal M$ and $\cal S$ are correction factors, respectively, to
relate $B_\varpi^+$ to $-F$ and $z_0\partial B_z/\partial \varpi$ to
$[(1-f_0^2)/(1+f_0^2)]\sigma^{-1}d \sigma/d\xi$ approximately at the
origin.  Although a proper treatment would require a procedure of
singular perturbation theory of the type described in \S 3.4, we
bypass such an involved treatment by the simple act of choosing ${\cal
M}$ and ${\cal S}$ to be the same thickness modification factor that
we used to regularize $F$,
\be
{\cal M} = {\xi\over \sqrt{\xi^2+\zeta^2(\xi)}} = {\cal S}.
\ee

\begin{figure}
\figurenum{4}
{\centerline{\epsscale{0.90} \plotone{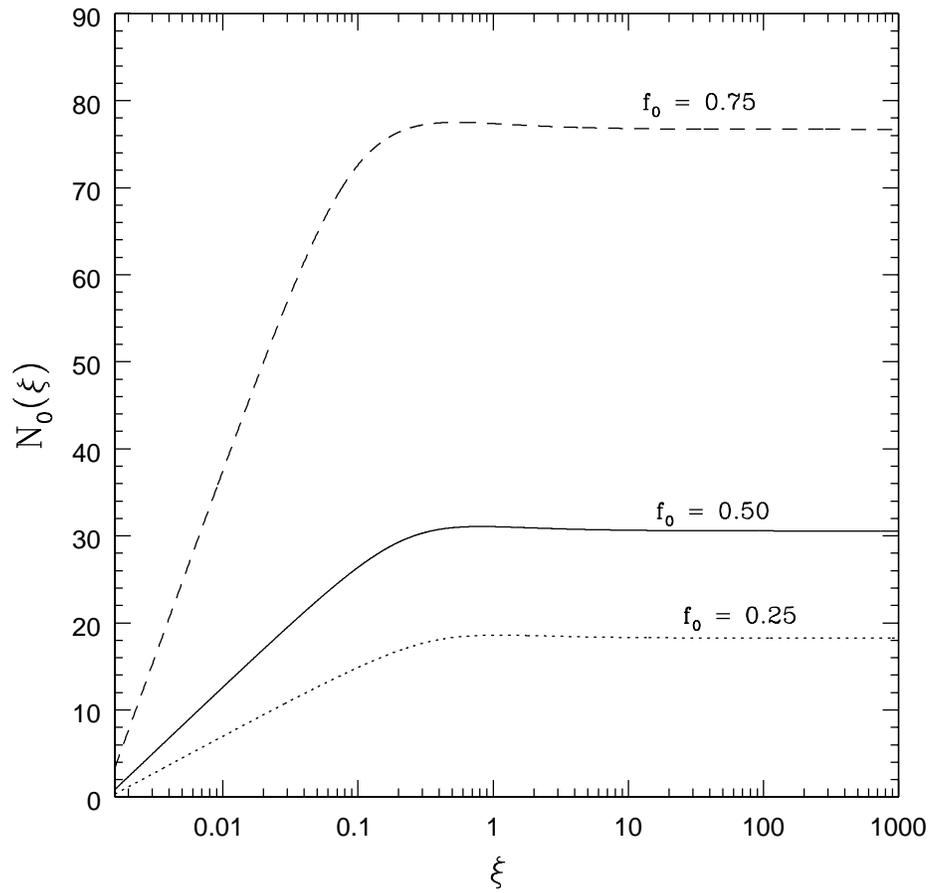} }} 
\figcaption{ The integral $N_0(\xi)$ as a function of
$\xi$.  The three curves correspond to different values of the zeroeth
order flux to mass ratio: $f_0$ = 0.75 (dashes), $f_0$ = 0.50 (solid),
and $f_0$ = 0.25 (dots).} 
\label{fig:nintegral} 
\end{figure}

With this procedure, ${\cal N}_{\rm L}$ goes to a constant at $\xi =
0$ rather than diverging as $1/\xi$ near the origin.  With ${\cal
N}_{\rm L}$ replacing ${\cal N}$, we evaluate the integral
\be
N_0(\xi) \equiv \int_0^\xi {\cal N}_L (\xi) \, d\xi .
\ee
The resulting functions for $N_0(\xi)$ are shown in Figure
\ref{fig:nintegral} for the zeroeth order flux ratios $f_0$ = 0.25,
0.50, and 0.75. We note that the numerical values for $N_0(1)$ are
18.6, 31.1, and 77.4 for the cases $f_0$ = 0.25, 0.50, 0.75, 
respectively. Furthermore, the values of $N_0(\infty)$ are nearly
equal to those of $N_0(1)$, with differences of only $\sim2\%$.  
We denote the value of $N_0(1)$ as a function of $f_0$ by the symbol
$I(f_0)$ and we rewrite equation (\ref{eigenvalue}) to obtain the 
following explicit relationship between $\epsilon$ and $f_0$:
\be
\epsilon = {(1-f_0) \sqrt{1+2f_0^2} \over f_0^3 I(f_0)}. 
\label{epsf0}
\ee
Note that the correlation of $\epsilon$ with $f_0$ is extremely
sensitive. Specifically, in order to obtain flux to mass ratios $f_0$
= 0.75, 0.50, and 0.25, the required values of $\epsilon$ are 0.0112, 
0.157, and 2.73, respectively.  We cannot obtain a consistent
approximation with $\epsilon \ll 1$ in this formulation when $f_0$
becomes too low.  Conversely, we may say that for standard values of
$\epsilon$ ($\approx 0.18$), runaway core condensation occurs
typically when $f_0$ reaches 0.5, as indicated by simulations
performed by many different groups and under very different
assumptions (e.g., Nakano 1979, Lizano \& Shu 1989, Basu \&
Mouschovias 1994).  Values of $f_0$ appreciably smaller than 0.5
requires anomalously large values of $\epsilon$, under which the
condensation problem becomes highly dynamic and can hardly be
considered diffusive.  Such cases, if they exist, need alternative
treatments to the one given here. As indicated by equation
(\ref{epsf0}), the limiting cases $f_0 = 0$ (spherical unmagnetized
core) and $f_0 = 1$ (completely flattened, critically magnetized core)
are special in that the first cannot sensibly attach onto a critically
magnetized envelope unless $\epsilon$ is infinite, whereas the second
loses flux from its central regions and becomes inconsistent with the
assumption that $f_0 =1$, the same value as the common envelope,
unless $\epsilon$ is zero.

Although our derivation of the important equation (\ref{epsf0}) was
carried out for a very specific model, we believe that the result is
robust.  Indeed, the result is almost given by dimensional analysis,
except we shall do it through a dimensionless argument. Consider the
vector induction equation with ambipolar diffusion as it is given by
equation (27.12) of Shu (1992), 
\be
{\partial {\bf B}\over \partial t} + \nabla \times 
({\bf B}\times {\bf u}) = \nabla \times \left\{
{{\bf B}\over 4\pi \gamma {\cal C}_{\rm local}\rho^{3/2}}
\times [ {\bf B}\times (\nabla \times {\bf B})]\right\}.
\label{vectorambi}
\ee
where we have specialized to the ionization law $\rho_{\rm i} = 
{\cal C}_{\rm local}\rho^{1/2}$ (Appendix B).  When applied to the 
core-formation problem, the right-hand side (RHS) of equation
(\ref{vectorambi}) is proportional to three powers of ${\bf B}$ and
inversely to the product $\gamma {\cal C}_{\rm local}$. In
dimensionless form, the RHS is proportional to $\epsilon f_0^3$ in the
central regions of a condensing core.  The left-hand side (LHS)
measures the distance (in time or distance divided by velocity) that
the magnetic field in the core has to travel to get from some starting
envelope or boundary value to the central value.  In proper
dimensionless form, this distance is $1-f_0$. This task is
accomplished at the rate $\propto \epsilon f_0^3$ on the RHS;
therefore, when the LHS equals the RHS, we have $(1-f_0) \propto
\epsilon f_0^3$.  An order unity quantity on the LHS cannot equal an
order $\epsilon$ quantity on the RHS unless there is a relatively
large proportionality factor on the RHS. Although this factor might
depend on $f_0$ (because of the specifics of the model), the
dependence should be fairly slow since the major dependences should be
captured by our scaling arguments.  The proportionality factor
$I(f_0)/\sqrt{1+2f_0^2}$ in equation (\ref{epsf0}) has precisely these
two qualities: (1) it is relatively large, and (2) it is relatively
constant as a function of $f_0$.  Thus, apart from minor quibbles
about exactly what function $I(f_0)/\sqrt{1+2f_0^2}$ should be, the
relationship derived as equation (\ref{epsf0}) is insensitive to the
details of geometry, or whether ambipolar diffusion is primarily
driven by pressure gradients or magnetic tension, etc.

The physics behind why it is difficult to drive the central
flux-to-mass ratio $f_0$ to low values is now obvious.  As ambipolar
diffusion occurs and $|{\bf B}|$ decreases, the rate of diffusion,
proportional not only to $\epsilon$ but also to $|{\bf B}|^3$, slows
down appreciably. It thus becomes increasingly difficult to make
$|{\bf B}|$, relative to $\Sigma$, which is the proper comparison
field for $|{\bf B}|$, even smaller.  As a result, many condensing
cores get stuck around $f_0 \approx 0.5$ before gravitational
instability takes over and the ratio $|{\bf B}|/\Sigma \propto f
\propto \lambda^{-1}$ is swept into regions close to the origin for
further adventures in the exciting process called star formation
(Mouschovias 1976, Shu et al. 2007).

\begin{figure}
\figurenum{5}
{\centerline{\epsscale{0.90} \plotone{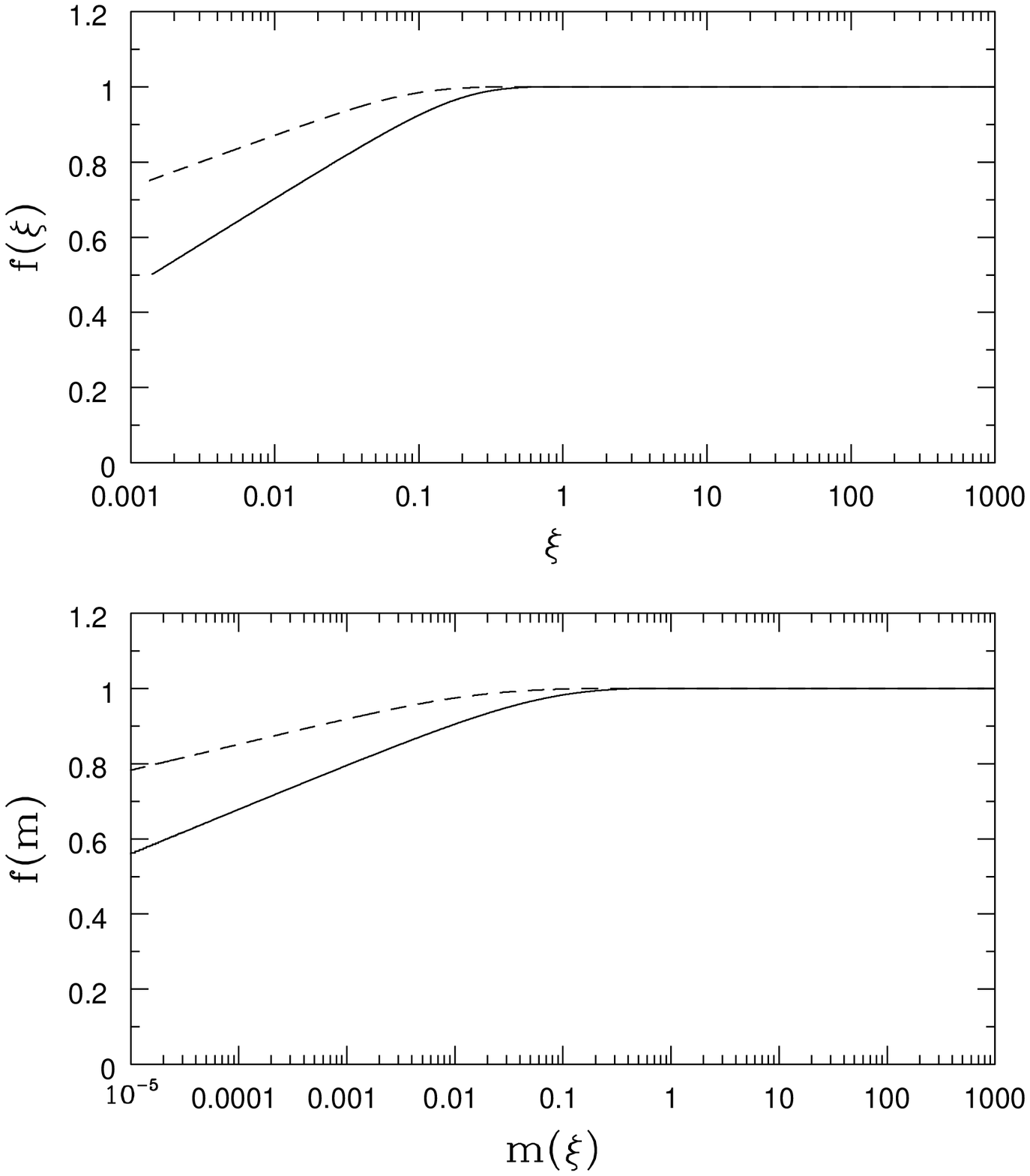} }} 
\figcaption{ Flux to mass ratios. Top panel shows the profiles $f(\xi)$ 
for $f_0$ = 0.50 (solid curve) and $f_0$ = 0.75 (dashed curve). Bottom panel 
shows the flux to mass ratio $f(m)$ as a function of enclosed mass $m(\xi)$ 
for the same cases, i.e., $f_0$ = 0.50 (solid curve) and $f_0$ = 0.75 (dashes). } 
\label{fig:fluxprofile} 
\end{figure}

To see graphically the formal flux-to-mass profiles implied by our
models, we rewrite equation (\ref{gensolnf}) in the present notation as 
\be
f(\xi,\tau) = f(1,\tau_1)-{(1-f_0)\over N_0(1)}\left[ N_0(1)-N_0(\xi)\right].
\label{finalf}
\ee
Before runaway condensation occurs, there is not much fluid motion (if
we ignore the presence of turbulence), and every region can be
connected to a nearly static common envelope by characteristics from
the past, so $f(1,\tau_1) = 1$. For $\xi > 1$, $N_0(\xi)$ is
essentially equal to $N_0(1)$ and we have $f(\xi,\tau) = 1$, which
identifies such regions as corresponding to the common envelope.  For
$\xi < 1$, $N_0(\xi) < N_0(1)$ according to Figure \ref{fig:nintegral}, 
and $f(\xi,\tau)$ has a value less than 1, i.e., the common envelope
makes a transition to a core region undergoing ambipolar diffusion
with magnetic flux leaking from the $\xi < 1$ core into the common
envelope $\xi > 1$.  In particular, for $\xi = 0$, $N_0(\xi) = 0$, so
the central flux-to-mass value of the core is $f(0, \tau) = f_0$.  
The top panel of Figure \ref{fig:fluxprofile} shows the flux profile
$f(\xi) = 1-[(1-f_0)/N_0(1)][N_0(1)-N_0(\xi)]$ before the onset of
runaway condensation for the realistic cases of $f_0 = 0.75$ and $f_0
= 0.50$.  Because the function $N_0(\xi)$ is computed assuming
values of $\sigma (\xi)$ and $v(\xi)$ applicable during runaway
condensation (see Fig. \ref{fig:nintegral}), this profile should
really be considered appropriate only for the specific time $\tau = 1$
demarcating the {\it transition} from quasi-static evolution by
ambipolar diffusion to runaway core condensation.

After runaway condensation occurs, we can no longer set the advective
contribution $f(1,\tau_1)$ equal to 1 because $\xi > 1$ at $\tau \ll
1$ lies in the {\it future} from $\xi =1$ at $\tau_1 \sim 1$ where
ambipolar diffusion has already occurred to modify $f$ from the value
of unity.  Instead, $f(1,\tau_1)$ becomes the dominant {\it variable}
term. In comparison, the {\it ongoing} diffusion term proportional to
$\hat \epsilon$ becomes increasingly negligible as the difference
$N_0(1)-N_0(\xi)$ vanishes because all relevant values of $\varpi$
correspond to $\xi = \varpi/\Theta^{1/2}at_0\tau > 1$.  The physical
meaning of this result is that approximate field freezing applies
during the phases of runaway condensation followed by true
gravitational collapse, and the flux-to-mass ratio plotted as a
function of interior mass becomes fixed in the subsequent evolution.
Only when very high densities are reached is the assumption of field
freezing again violated, perhaps involving the formation of a
circumstellar disk if we had included the effects of core rotation
(see the discussion in \S 7).

The bottom panel of Figure \ref{fig:fluxprofile} plots the flux
profile $f$ versus the reduced and scaled enclosed mass $m(\xi)$ for
the same cases depicted in top panel. This plot shows that the central
value $f_0$ is a substantial under-estimate of the average
flux-to-mass ratio of the entire core, i.e., most of the core has a
flux-to-mass ratio closer to unity.  Correction for the under-estimate
should act to reduce the head-start velocities of real molecular cloud
cores in comparison with the values given in Table 1.  In particular,
we can expect extended contraction of the magnitude $-v_\infty$ only
where and when $f$ is still climbing to unity.  Application of this
rule to Figure \ref{fig:fluxprofile} indicates that extended
contraction might be observed, perhaps, to about 1/3 of the distance
to the outer core boundary (about 1/3 of the enclosed core mass). As
mentioned earlier, these estimates can be made more rigorous using
singular perturbation theory with multiple length scales, where the
region discussed in \S 5 is treated by ``intermediate asymptotics''
with a scale between the large ones of the common envelope and the
small ones of the runaway condensation that produces a gravomagneto
catastrophe as $t \rightarrow 0^-$.  Although we have not performed
such an improved analysis, we hope that the naive treatment of this
paper elucidates the {\it physical} basis of the phenomenon of
gravomagneto catastrophe.

In any case, at the moment of gravomagneto catastrophe, the enclosed
mass in physical units inside $\varpi = \varpi_{\rm ce}$ is given by
\be
M_{\rm core} = {A\Theta_0 a^2\over (1-f_0^2)G} \varpi_{\rm ce} = 
A\left( {1 + 2 f_0^2 \over 1 - f_0^4}\right) \mbench \, , 
\label{coremass} 
\ee
where the second equality uses the definition of $\Theta_0$ and
defines a benchmark mass scale $\mbench = a^2 \varpi_{\rm ce} / G$.
For typical values of $a$ = 0.20 km/s and $\varpi_{\rm ce}$ = 0.2 pc,
for example, $\mbench \approx 1.9 M_\odot$.  The range of $f_0$ is
limited because values of $f_0 < 0.3$ imply values of $\epsilon$
greater than unity. Over the range from 0.3 to (say) 0.9, (roughly, 
$1 > \epsilon > 0.002$), the core masses implied by equation
(\ref{coremass}) vary from $2.5 \mbench$ to $11 \mbench$. Note that
this range in mass scale, a factor of 4.4, is smaller than the
observed range of stellar masses. However, the values of $a^2$ and
$\varpi_{\rm ce}$ that specify the mass scale $\mbench$ can also vary,
and the distribution of these parameter values will add additional
width to the resulting distribution of core masses. Moreover, final
stellar masses can be appreciably smaller than the core masses at the 
beginning of dynamical collapse because of various inefficiencies in
the actual star-formation process (such as binary formation and
magnetocentrifugally driven winds leading to bipolar outflows). These
variations will also add width to the distribution of stellar masses
(see Adams \& Fatuzzo 1996 for greater mathematical detail). Taken as
a whole, a strength of ambipolar diffusion as a core-formation
mechanism is that, given plausible variations of $a^2$ and
$\varpi_{\rm ce}$, it is capable of producing a core-mass distribution
wide enough, when the pivotal state is reached, to span the likely
pre-collapse states for making brown dwarfs to high-mass stars.

\section{SHEET COLLAPSE SOLUTION} 

The analysis presented thus far accounts for the formation of
centrally condensed molecular cloud cores through the process of
ambipolar diffusion. This phase corresponds to negative times ($t <
0$) and can be smoothly matched onto collapse solutions at positive
times ($t > 0$).  Although the collapse portion of the problem has
been considered previously (e.g., Shu 1977, Hunter 1977, Galli \& Shu
1993, Li \& Shu 1997, Krasnopolsky \& K{\"o}nigl 2002, Shu et al.
2004, Fatuzzo et al. 2004), here we present a brief re-examination of
the problem, and find the particular solution that matches onto the
solution to the pre-catastrophe problem found in \S 4 and \S 5.

The equations of motion for collapse are the continuity equation
(\ref{continuityeq}) and the force equation (\ref{force}). For
simplicity, in this treatment we assume flux freezing during dynamical
collapse until very small scales are reached (see Galli \& Shu 1993,
Galli et al. 2006, and Shu et al. 2006) so that $g + \ell = (1 -
f_0^2) g$. Also to keep the discussion uncomplicated, and because
there is now a true physical monopole (the protostar) to keep the
inflowing material spatially flat, we consider the sheet monopole
limit for the gravitational force (see \S 4).  Here the relevant
similarity transformation has the form
\be 
x = {\varpi \over at} \, , \qquad 
\Sigma (\varpi, t) = {a \over 2\pi G t} \tilde \sigma(x) \, , \qquad 
{\rm and} \qquad u (\varpi, t) = a \tilde v(x) \, . 
\ee
After some algebra, the dimensionless self-similar form of 
the equations of the motion become 
\be 
\tilde \sigma {d \tilde v \over dx} + (\tilde v - x) 
{d \tilde \sigma \over dx} = \tilde \sigma {(x - \tilde v) \over x} \, , 
\ee
and 
\be
(\tilde v - x) {d \tilde v \over dx} + {\Theta_0 \over \tilde \sigma} 
{d \tilde \sigma \over dx} = (1 - f_0^2) 
\tilde \sigma { (\tilde v - x) \over x } \, , 
\ee 
where $\Theta_0$ and $f_0$ have the same meaning as before 
and are taken to be constants.

Next we apply the adopted scaling transformation 
\be 
\xi = {x \over \sqrt{\Theta_0} } \, , \qquad 
v = {\tilde v(x) \over \sqrt{\Theta_0} } \, , \qquad 
{\rm and} \qquad 
\sigma = \tilde \sigma (x) {(1 - f_0^2) \over \sqrt{\Theta_0} } \, . 
\ee 
The equations of motion then take the forms 
\be
\discrim {d { v} \over d\xi} = 
{\xi - { v} \over \xi} \left[ 
{ \sigma} (\xi - { v}) - 1 \right]  \, , 
\ee
\be
\discrim {d { \sigma} \over d \xi} = 
{ \sigma} {(\xi - { v}) \over \xi}
\left[ { \sigma} - (\xi - { v}) \right]  \, , 
\ee
where the discriminant is now given by 
\be
\discrim = (\xi - { v})^2 - 1 \, . 
\ee 

In the outer limit $\xi \to \infty$, the equations of 
motion allow the asymptotic forms 
\be 
{ \sigma} = {A \over \xi} 
\qquad {\rm and} \qquad 
{ v} = v_\infty + {(1 - A) \over \xi} \,  . 
\ee
In order for the collapse solution to match onto the pre-catastrophe
solution for $t < 0$, the constants $A$ and $v_\infty$ must be the
same of those of \S 4.3. Notice also that the sign of the correction
term in the velocity field is different for the two cases, as it
should be: The limit $\xi \to \infty$ corresponds to $t \to 0$ from
either side of zero, where the solutions must match, and where both
solutions have velocity $v_\infty$ (which is negative, since the core
is contracting).  For large but finite $\xi$, and negative times, the
condensation solution has a positive correction to the velocity, so
that the velocity is smaller in magnitude, i.e., it hasn't reached its
full head-start speed that it will at the moment of gravomagneto
catastrophe $t = 0$.  For large but finite $\xi$, and positive times,
the collapse solution has a negative correction to the velocity,
indicating that the fluid is speeding up as it collapses.

In the inner limit $\xi \to 0$, the scaled equations 
of motion imply that the solutions have the form of a 
free-fall collapse flow, i.e., 
\be 
{ \sigma} = \left( {\mchat \over 2 \xi} \right)^{1/2} 
\qquad {\rm and} \qquad 
{ v} = - \left( {2 \mchat \over \xi} \right)^{1/2} \, , 
\ee
where 
\be 
\mchat = \lim_{\xi \rightarrow 0} m(\xi) = {\rm constant} \, . 
\ee

Finding the constant $\mchat$ is the most important result of the
numerical procedure, since the rest of the solution is then specified
by the similarity transformation. In particular, the dimensional
mass-infall rate $\dot M$ is given by
\be
{\dot M} = {a^3 \over G}\tilde m_0 = 
{(\Theta_0^{1/2}a)^3 \over G(1-f_0^2)}\mchat .
\label{mdotphys} 
\ee 
Relative to the standard formula, the scaling of the final expression
has the following mnemonic: (1) the relevant velocity to be cubed is
the magnetosonic speed $\Theta_0^{1/2}a$, and (2) the relevant
gravitational constant is the magnetically diluted value $(1-f_0^2)G$.

The solution for $m_0$ is specified by the pair of constants $(A,
v_\infty)$ that determine how the collapse solution for $t > 0$
matches onto the condensation solution for $t < 0$. Furthermore, all
viable pairs of boundary values $(A, v_\infty)$ correspond to states
that are overdense ($A > 1$) and/or with finite head-start velocity
$-v_\infty > 0$. These solutions thus correspond to the ``outer''
solutions in the nomenclature of Shu (1977) or the generalization to
include nonzero starting velocities (Fatuzzo et al. 2004). In any
case, for these ``outer'' solutions the flow does not go through a
critical point. As a result, one can directly integrate the equations
of motion from asymptotically large $\xi$ (where the solution matches
onto those of the previous section) down to small $\xi \ll 1$ to
determine the constant $\mchat$.

\begin{figure}
\figurenum{6}
{\centerline{\epsscale{0.90} \plotone{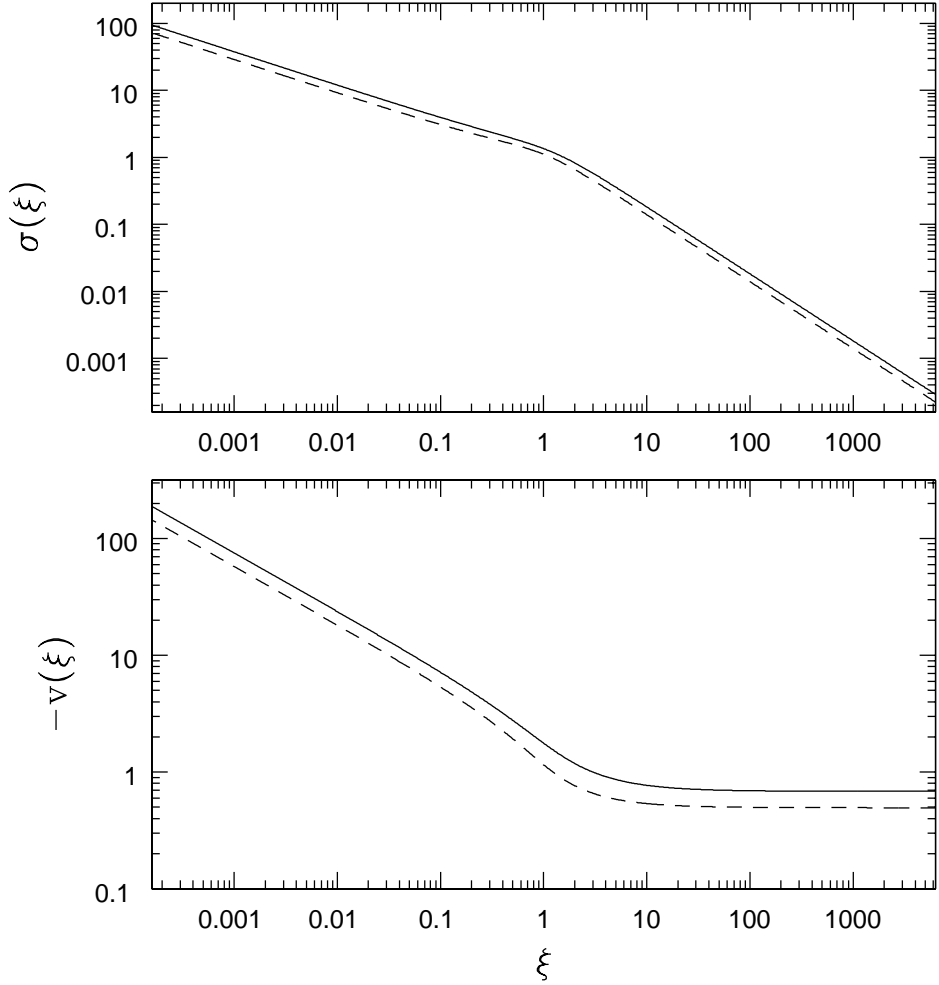} }} 
\figcaption{ Reduced fields for collapse solution in the sheet monopole 
(dashed curves) and softened monopole (solid curves) approximations. 
Upper panel shows the reduced and scaled density field $\sigma (\xi)$.
The lower panel shows the solution for the reduced and scaled velocity
field $v (\xi)$. The initial conditions for collapse are taken to be
those predicted from the condensation calculation (see
Fig. \ref{fig:monopole}). }
\label{fig:monocollapse} 
\end{figure}

For the sheet monopole solution to the $t < 0$ evolution, the boundary
values are $(A, v_\infty) = (1.401, -0.4952)$. Using these starting
conditions, the resulting collapse solution for $t > 0$ is shown as
the dashed curves in Figure \ref{fig:monocollapse}. For this case, the
inner constant $\mchat = 1.670$, about 70 percent larger than the
coefficient found by Shu (1977) for the collapse of the critically
stable, singular isothermal sphere, $\mchat$ = 0.975.  For the
softened monopole solution to the $t < 0$ evolution, the boundary
values are $(A, v_\infty) = (1.83, -0.684)$.  For this case, the
mass-infall constant for $t > 0$ is $\mchat = 2.85$, and the resulting
solution is shown as the solid curves in Figure
\ref{fig:monocollapse}.

We note that the collapse solution presented here is somewhat
idealized, even within the class of possible self-similar solutions.
The collapse flows shown in Figure \ref{fig:monocollapse} are
calculated from the sheet monopole and softened monopole
approximations for the gravitational field. For cases that include a
full calculation of the perturbational gravity and no initial inward
velocities (e.g., Li \& Shu 1997, Krasnopolsky \& K{\"o}nigl 2002), a
shock front develops just outside the infall region.  Except for the
region near the shock, which includes the transition between the inner
collapsing flow and the outer quasi-static region, the solutions with
and without shocks are qualitatively and quantitatively similar.
Specifically, the collapse solutions with monopole gravity (and no
shock front) result in a reduced point mass $m_0 \approx 1.3$, whereas
the case of full gravity solutions (with a shock front) result in $m_0
\approx 1.05$ (Li \& Shu 1997).  In the case considered here, however,
the $t=0$ configurations (at the end of the ambipolar condensation
phase and the start of the collapse phase) have non-zero inward
velocities which act to eliminate the critical points in the flow
(e.g., Fatuzzo et al. 2004), so that we do not expect shocks near the
head of the expansion wave to play a significant role in the collapse.

This treatment also neglects the effects of rotation on collapse. The
solutions found here thus represent the outer portion of the collapse
flow, and must be matched onto inner solutions that include rotation
(Cassen \& Moosman 1981, Terebey et al. 1984). When the inner portion
of the outer region approaches ballistic (pressure-free) conditions,
this matching can be done seamlessly (Shu 1977; Li \& Shu 1997;
Fatuzzo et al. 2004). For collapse flows that include magnetic fields,
however, the roles of magnetic braking and magnetorotational
instability (MRI) can be important (see Allen et al. 2003; Galli et
al. 2006; Shu et al. 2006, 2007).  The calculations of this paper show
that if field freezing strictly holds for the collapse phase $t > 0$,
then the value of $\lambda_0$ brought into the star plus disk would be
typically $\sim$2, which could prevent disk formation by magnetic
braking; this result has also been found in numerical simulations
(e.g., Fromang, Hennebelle, \& Teyssier 2006, Price \& Bate 2007).
How circumstellar disks form and evolve thus remains an open question,
although it appears likely that global MRI in a context of nonzero net
flux will play a major role (Shu et al. 2007).
 
\section{ILLUSTRATIVE NUMERICAL EXAMPLE} 

\begin{figure}
\figurenum{7}
{\centerline{\epsscale{0.90} \plotone{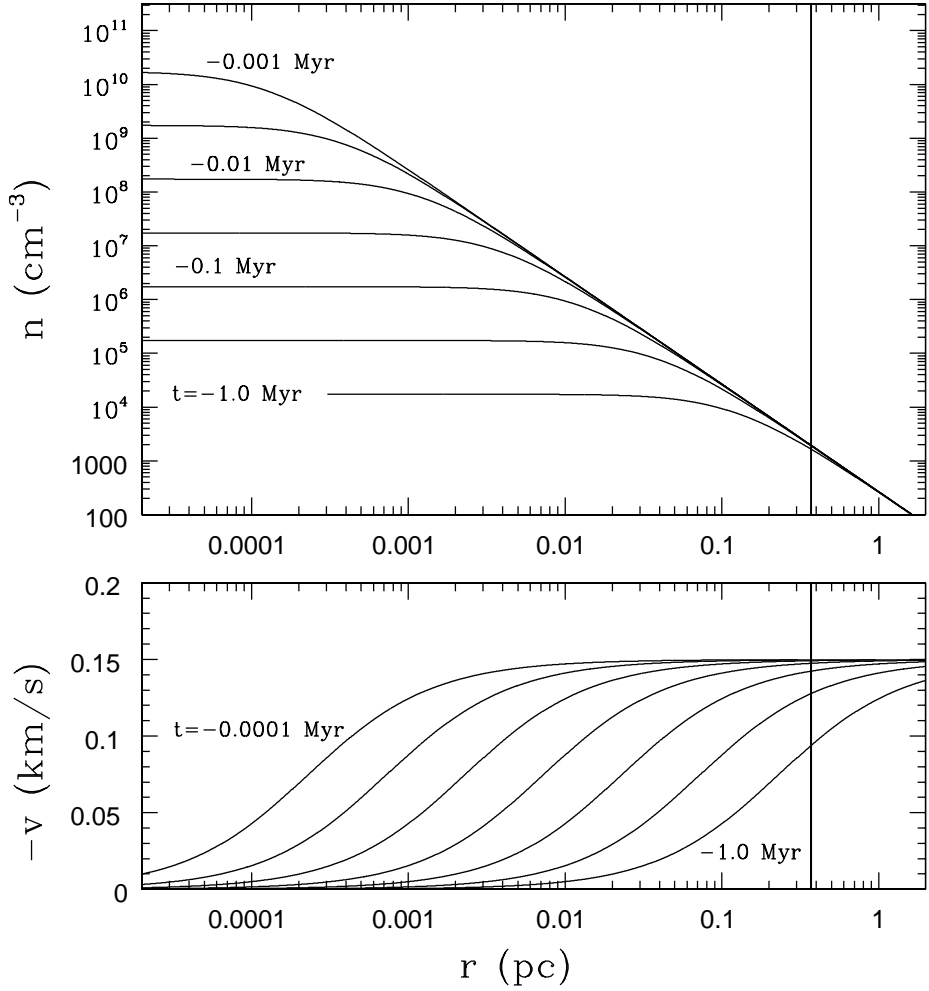} }} 
\figcaption{ Physical solutions for the number density and velocity of
a molecular cloud core during its condensation (formation) epoch for
varying times before the moment of gravomagneto catastrophe (which
occurs at $t$ = 0).  The displayed time levels are spaced
logarithmically, e.g., $-0.316$ Myr separates the curves labeled
$-1.0$ Myr and $-0.1$ Myr in the top panel.  The vertical solid lines
mark the location of the outer core boundary, if it were set by the
condition that the density falls to a benchmark value $n$ = 1000
cm$^{-3}$, where a core would join onto the background molecular
cloud. The displayed velocity field is not probably trustworthy when
one has reached about 1/3 of the distance to the outer core boundary. } 
\label{fig:condensephys} 
\end{figure}

\begin{figure}
\figurenum{8}
{\centerline{\epsscale{0.90} \plotone{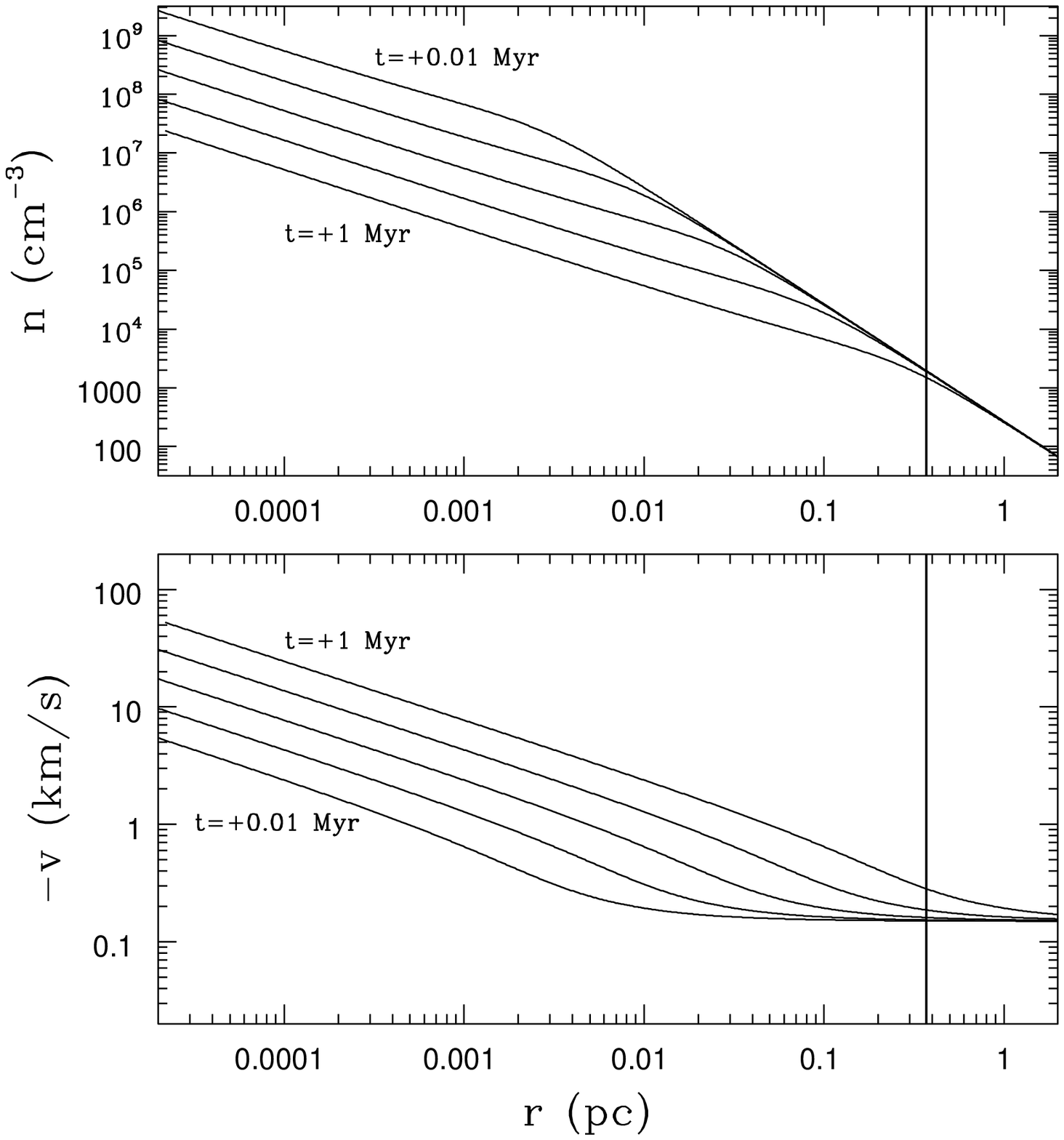} }} 
\figcaption{ Physical solutions for the number density and velocity of
a molecular cloud core during its collapse epoch for varying times
$t>0$ after the moment of catastrophe.  The five curves in each panel
correspond to logarithmically spaced time intervals, i.e., the times
$+0.0316$ Myr, $+0.1$ Myr, $+0.316$ Myr lie between the displayed
times of $+0.01$ Myr and $1$ Myr.  The vertical solid lines mark the
location of the outer core boundary, if it were set by the condition
that the density falls to a benchmark value of $n$ = 1000 cm$^{-3}$,
where a core would join onto the background molecular cloud. The
displayed region of extended infall is probably not probably
trustworthy when one has reached about 1/3 of the distance to the
outer core boundary. }
\label{fig:collapsephys} 
\end{figure}

To give astronomical context to the semi-analytic results of this
paper, we next plot the evolution given by the softened monople
condensation and collapse solutions in dimensional form for the case
$f_0 = 0.5$ and $a = 0.2$ km s$^{-1}$.  In Figures
\ref{fig:condensephys} and \ref{fig:collapsephys} we show the
equatorial volume density $\rho(\varpi,0,t) \equiv \Sigma/2z_0$,
plotted here as the number density $n = \rho/2.3 m_{\rm H}$, and the
equatorial inflow velocity $-u(\varpi,t)$ as functions of $\varpi$ and
$t$.  Figure \ref{fig:condensephys} shows the time evolution for
negative times $t < 0$ (starting from $t = -1.0$ Myr), whereas Figure
\ref{fig:collapsephys} shows the time evolution for positive times $t
> 0$ (out to $t = +1.0$ Myr). Notice that the evolution of the runaway
condensation phase for $t < 0$, as shown by Figure
\ref{fig:condensephys}, compares well with the calculations of
ambipolar diffusion carried out numerically by Basu \& Mouschovias
(1994).

The important thing to carry away from Figure \ref{fig:condensephys}
is that condensing cores are not observable in dense-gas tracers such
as NH$_3$, which requires $n > 3\times 10^4$ cm$^{-3}$ for excitation,
until the cores are within several hundred thousand years of gravomagneto
catastrophe.  If ``cores'' are defined as such by whether they are
observable in dense-gas tracers, then their ``lifetimes'' will be
comparable to the lifetimes of embedded protostars, also measured in
the several hundreds of thousands of years.  This numerical
coincidence results in roughly equal numbers for ``starless cores''
and ``cores with embedded stars,'' with considerable scatter depending
on the value of $\epsilon$ in the region being studied.  Similar
statistics given by observers, plus the finding that the surface
density profiles of cores are flat in their central parts, have led to
the mistaken criticism that ambipolar diffusion seems to work too
slowly to account for the observations (e.g., Andr\'e et al. 1996,
Ward-Thompson et al. 1999).  Runaway core condensation, the phase
depicted in Figure \ref{fig:condensephys}, does not take long, but it
is just the last stage of the ambipolar diffusion process.  Indeed, it
is a stage where not much ambipolar diffusion is still going on, with
the central flux-to-mass $f_0$ being ``frozen'' in value.  Prior to
this stage, there were slower stages of evolution, lasting maybe an
order of magnitude longer than $10^6$ yr (although it is hard to
specify when to start the clock for this less definite problem),
where ambipolar diffusion {\it did work} to get the central regions
into the runaway state (see Fig. \ref{fig:fluxprofile}).  These stages
are not well studied by the similarity methods of this paper but have
been amply treated by many numerical simulations (e.g., Nakano 1979,
Lizano \& Shu 1989, Basu \& Mouschovias 1994, Desch \& Mouschovias
2001), and they occupy the bulk of the evolution time starting from
arbitrarily chosen ``initial'' conditions.

The fact that the inflow velocities of ``extended contraction'' are
observed to be a significant fraction of the sound speed $a$ indicates
that once runaway condensation commences, the process is fairly rapid.
Nevertheless, if the extended inflow velocities typically reach half the
magnetosonic speed, the core is roughly within 75\% of being
``magnetohydrostatic.'' As a result, the oft-repeated statement that
``star formation is a dynamical process'' is an illusion when applied
to $t < 0$, and it is a tautology when applied to $t > 0$. Observations
of extended converging flows thus do not support the view of star
formation as a turbulent dynamic process, nor do they refute the
theory of ambipolar diffusion as the core formation mechanism. Indeed,
as we argue in \S 8, the magnitude of the observed inflow velocity is
fully consistent with the predictions of this paper, and thus provides
a strong argument that molecular-cloud core-formation, at least in
isolated regions of relatively low total mass, is due to ambipolar
diffusion.

For the collapse phase ($t > 0$), Figure \ref{fig:collapsephys} shows
that the fluid fields display the usual forms, as studied in many
previous treatments (e.g, Shu 1977, Hunter 1977, Galli \& Shu 1993,
Allen et al. 2003, Fatuzzo et al. 2004). In particular, as the
collapse proceeds, the densities at a given radius $\varpi$ become
increasingly {\it smaller} than in the pivotal state before it. This
behavior is a consequence of the solution belonging to the family of
{\it expansion-wave} collapse solutions (Shu 1977), with the material
missing from the inner core collecting in a point-like object (the
forming star) at the center of the collapse flow.  In the new case
considered here, the collapse solution at positive times matches
smoothly onto the condensation solution of negative times. Not only
does this generalization provide a self-contained picture of core
formation and subsequent collapse, but it also shows that the starting
state for collapse has a nonzero inward velocity, which results in a
somewhat larger mass infall rate. The trend of decreasing density at a
given radius $\varpi$ is thus more extensive than in the case of a
magnetohydrostatic starting state, because the ``piston'' is pulled in
by the head-start velocities more rapidly and in a more continuous
manner.

\section{DISCUSSION AND CONCLUDING REMARKS}

In this paper, we have presented a picture of molecular cloud
core-formation that envisages a process of the slow leakage of
magnetic flux from a dense pocket of gas and dust via ambipolar
diffusion. This process continues until the central regions acquire a
mass to flux to ratio $\lambda_0 = f_0^{-1}$ that is too high for
continued quasi-hydrostatic support against the self-gravity of the
core, and the core develops a runaway central density with a power-law
profile $\Sigma \propto \varpi^{-1}$ or $\rho \propto r^{-2}$ at the
moment of gravomagneto catastrophe $t= 0$.  One of the key findings of
this paper is that $f_0$ is given by the root of equation
(\ref{epsf0}): 
\be
{(1-f_0)\over f_0^3 K(f_0)} = \epsilon,
\ee
where $K(f_0) \equiv I(f_0)/\sqrt{1+2f_0^2})$ $\sim 25$ for values
of $f_0$ of physical interest, and $\epsilon$ is the dimensionless
rate of ambipolar diffusion given by equation (\ref{epsdef}).  This
equation shows that it is difficult for ambipolar diffusion to produce
regions with $f_0 < 0.3$ without an anomalously high effective rate
coefficient $\epsilon > 1$.  A more typical outcome is probably 
$f_0 \sim 0.5$, although values of $f_0$ close to unity might be
sustainable in regions with high ionization rates (which lowers
$\epsilon$).  Such regions will be characterized by high surface
densities $\Sigma$ $\propto (1-f_0^2)^{-1}$, and therefore large
visual extinctions, as well as large core masses, and could be one
ingredient to the cores that make massive stars.  Another ingredient
could be high gas temperatures or high levels of turbulence.
 
During the epoch of core formation, $t < 0$, extended regions of
contraction develop, and extend to perhaps one third of the way to the
effective boundary where the core joins onto a common envelope of the
dense clump that surrounds it (see \S 5.3 and Fig. 5). Significantly,
the contraction velocities are always beneath the magnetosonic value.
A transition to fully dynamic flow is made for $t > 0$ that
corresponds to an inside-out collapse solution, but with a
sub-magnetosonic head-start velocity and an over-dense outer envelope
given to the region by the ambipolar diffusion process in the previous
epoch $t < 0$. The main feature of the $t > 0$ collapse solution is
self-similar infall onto a growing protostar at a mass infall rate
$m_0 [\Theta_0^{1/2} a]^3/(1-f_0^2)G$, where the dimensionless
coefficient $m_0$ is two or three times larger than the ``standard'' 
value of 0.975 (resulting from purely hydrostatic states at $t=0$). 
In some sense, this new solution combines the most attractive features
of the self-similar solutions proposed in previous work (e.g., Larson
1969, Penston 1969, Shu 1977, Hunter 1977, Fatuzzo et al. 2004)
without bearing any of the unncessary baggage.

The models predict that the head-start velocities are correlated with
over-densities.  Compared to the static singular isothermal sphere
(SIS), the sub-magnetosonically inflowing parts of the core
characterized by $r^{-2}$ volume densities have total over-density
factors of $A(1+2f_0^2)/(1-f_0^4)$. For $f_0 = 0.5$ in the softened
monopole model where $A = 1.83$ (see Table 1), this amounts to an
overall factor of 3.  From a near-infrared extinction study of the
bipolar outflow source B335, Harvey et al. (2001) found the outer
portions of its associated core to have a $r^{-2}$ power-law behavior
for the inferred volume density, with an over-density relative to the
SIS of 3 to 5.  They interpreted their data in terms of an unstable
Bonnor-Ebert sphere.  Later, Harvey et al. (2003a) found the inner
portions of B335 to have a density profile consistent with a
$r^{-3/2}$ law, i.e., consistent with the detailed modeling of this
source as a classic example of inside-out collapse (Zhou et al. 1993,
Choi et al. 1995, Evans et al. 2005).  In a study of the starless
globule L694-2, which has strong evidence for inward motions, Harvey
et al. (2003b) find the outer portion of the core to have a $r^{-2.6}$
volume density profile, steeper than our model predictions, but with
an over-density factor in its central regions relative to Bonnor-Ebert
extrapolations of about 4.  While these authors speculate that the
effect might arise from the core being an prolate object viewed along
its long axis, a more satisfying and unified interpretation is that
B335 is a $t > 0$ post-catastrophe core with an embedded protostar
(and bipolar outflow), magnetized at $f_0\approx 0.5$; and that L694-2
is a $t < 0$ pre-catastrophe starless core, magnetized also at $f_0
\approx 0.5$.  Another indicator of the correctness of this
identification is that the predicted mass-infall rate ${\dot M} = m_0
[\Theta_0^{1/2} a]^3/(1-f_0^2)G$ is roughly consistent with measured
values in Class 0 sources (for examples with estimates at the
extremes, see Ohashi et al. 1997 and Furuyu et al. 2006), despite
earlier claims that Class 0 protostars would have much higher infall
rates (e.g., Henriksen et al. 1997).

The (near) self-similarity of the problem is a particularly attractive
feature of the process.  Given that the central portions of the core
are nearly isopedic, i.e., that $\lambda = 1/f$ is nearly a spatial
constant, self-similarity of the collapse solution for $t > 0$ (but
not too much greater!), or even for the runaway core-condensation
phase for $t < 0$ (but not too much less!) is perhaps not a surprising
outcome of nature's tendency to produce power-laws when solutions have
to span large dynamic ranges in space and time.

However, the reader could rightfully question
whether the $f$-profiles obtained in Figure \ref{fig:fluxprofile} are
not special to the application of self-similarity to a problem -- the
initial stages of the condensation of a cloud core by ambipolar
diffusion -- that has no good reason, beyond mathematical convenience,
to be self-similar.  After all, the contraction of typical molecular
densities in clumps of $\sim 10^3$ cm$^{-3}$ to early-stage cloud
cores with number densities $\sim 3\times 10^4$ cm$^{-3}$ can hardly
be characterized as spanning a very large range.  One might think that
the resulting $f$-profiles would show considerable variation depending
on the exact initial state being assumed and boundary conditions being
applied.  And so it must be with very detailed descriptions of such
early stages of cloud evolution.  But if one is pressed for more
global trends, there are only so many ways that a function $f$ can
monotonically go from unity at some large radius to some other value
$f_0$, typically 1/2, at some small radius.  And a self-similar
approach to getting such a profile is probably not any worse than some
other ad hoc prescription.  The important features of the picture are
not the details of the $f$-profiles, but the global view provided by
estimates of the relevant time scales, relationships between
$\epsilon$ and $f_0$, generic stages of the evolution, and the final
asymptotic convergence to self-similarity as the moment of
gravomagneto catastrophe is approached and passed.

In particular, the solutions presented in this paper are compatible
with full ambipolar-diffusion calculations that start with marginally
subcritical configurations which develop nearly isopedic central cores
(with $\lambda \approx$ constant) before proceeding on a path of
extended gravitational condensation that leads to gravomagneto
catastrophe (Nakano 1979, Lizano \& Shu 1989, Basu \& Mouschovias
1994).  It is particularly significant that both the predicted and
observed contraction velocities are sub-magnetosonic and arise from
the modest over-densities that are left behind in the contracting
cores as their magnetic support leaks to the common envelope.  Thus,
the observed head-start velocities are an indicator that some slow
process like ambipolar diffusion is at work producing molecular cloud
cores, rather than some more sudden process of the destruction of high
levels of non-thermal support, such as the dissipation of hypersonic
turbulence through shock waves.  The latter description may still
apply, however, in the crowded conditions that characterize high-mass
star-forming regions.

The most important lesson of this paper is that the details of the
ambipolar diffusion process control only the spatial extent of the
region of extended, sub-magnetosonic, contraction and the timing of
the runaway core-condensation that leads to the gravomagneto
catastrophe.  Provided the small parameter $\epsilon$ is not strictly
zero, gravomagneto catastrophe is the unavoidable fate of a
lightly-ionized, isolated, molecular-cloud core, as long as Lorentz
forces contribute to the support against its self-gravitation (see the
discussion of Lizano \& Shu 1989 concerning ``failed cores'').  Given
that the inner cores acquire nearly isopedic states with $\lambda =1/f
\approx$ constant, the resulting density and magnetic field profiles
of the resulting runaway condensation steepen into generic power laws
and are robust.

\acknowledgments 
This work was initiated during a sabbatical visit of FCA to UCSD; we
would like to thank the Physics Departments at both the University of
Michigan and the University of California, San Diego, for making this
collaboration possible. Discussions with Mike Cai were very helpful.
This work was supported through the University of Michigan by the
Michigan Center for Theoretical Physics; by NASA through the
Astrophysics Theory Program (NNG04GK56G0) and the Spitzer Space
Telescope Theoretical Research Program (1290776).

\appendix 
\section{Magnetic Forces in a Thin Disk}

In this Appendix, spurred by a correction pointed out by Mike Cai
(2007, private communication), we revisit the derivation given by Shu
\& Li (1997) for the magnetic forces in a thin disk.  The Lorentz
force per unit volume with only axisymmetric poloidal fields is given by
$$
{1\over 4\pi} (\nabla \times {\bf B})\times {\bf B} = 
{1\over 4\pi}\left( {\partial B_z\over \partial \varpi}-
{\partial B_\varpi \over \partial z}\right)
\left( B_z \hat {\bf e}_\varpi -B_\varpi \hat {\bf e}_z\right) .
\eqno{\rm (A1)}
$$
If we integrate over $z$ in a thin disk of effective thickness
$2z_0\ll \varpi$ with $B_\varpi^+ = -B_\varpi^-$ being, respectively,
the radial magnetic field at the upper and lower surface of the disk
and with $B_z$ being continuous across the midplane, the force per
unit area in the radial direction is given by
$$
-{B_z B_\varpi^+\over 2\pi}-{\partial \over \partial \varpi}
\left({B_z^2 z_0\over 4\pi}\right).
\eqno{\rm (A2)}
$$ 
We recognize the first term as what Shu \& Li refer to as the force
per unit area due to magnetic tension, but the second term is only the
negative radial gradient of the vertically integrated magnetic
pressure due to $B_z^2/8\pi$ and not $(B_z^2+B_\varpi^2)/8\pi$.  The
reason is that the so-called ``magnetic-tension term'' also contains a
small piece of the magnetic pressure, in fact, exactly the integral of
$B_\varpi^2/8\pi$ over the disk thickness.  Nevertheless, for
simplicity we shall continue to refer to the two terms as ``magnetic
tension'' and ``magnetic pressure.''

Vertical hydrostatic equilibrium for a magnetized isothermal gas in
its own vertical gravitational field requires
$$
-\rho {\partial U\over \partial z}-a^2
{\partial \rho\over \partial z}-{\partial \over \partial z}
\left( {B_\varpi^2\over 8\pi}\right) = 0,
\eqno{\rm (A3)}
$$
where the last term is the dominant term for the magnetic force per
unit volume in the $z$ direction according to equation (A1).  In the
above, $U$ is the self-gravitational potential of the gas and
satisfies the local Poisson's equation:
$$
{\partial^2U\over \partial z^2} = 4\pi G\rho .
\eqno{\rm(A4)}
$$
The substitution of the above into equation (A3) 
allows us to integrate once:
$$
{1\over 8\pi G}\left({\partial U\over \partial z}\right)^2
+a^2\rho +{B_\varpi^2\over 8\pi} = C(\varpi),
\eqno({\rm A5})
$$
where $C(\varpi)$ is a constant for fixed $\varpi$ (and diffusion
time).  We evaluate $C$ at the upper disk surface where $\partial
U/\partial z = 2\pi G \Sigma$, $\rho = 0$, and $B_\varpi =
B_\varpi^+$; and we do the same at the disk mid-plane where $\partial
U/\partial z = 0$, $\rho \equiv \Sigma/2z_0$ (being a definition of
$z_0$), and $B_\varpi = 0$.  Setting equal the two expressions for the
``constant'' total pressure, we obtain
$$
{a^2\Sigma \over 2z_0} = {\pi G\Sigma^2 \over 2}+{(B_\varpi^+)^2\over 8\pi}.
\eqno{\rm (A6)}
$$

For an isopedic singular isothermal disk, which is what the inner
parts of molecular cloud cores become at the moment of gravomagneto
catastrophe,
$$
B_\varpi^+ = B_z = {2\pi G^{1/2}\Sigma \over \lambda},
\eqno({\rm A7})
$$
with $\lambda$ equal to a constant. Equation (A5) now becomes the
second relation of equation (\ref{closure}).  In the same limit, we
have
$$
{B_z^2 z_0 \over 4\pi} =  \left({a^2\Sigma\over 1+\lambda^2}\right),
$$
which shows that the sum of gas pressure force in the radial direction
with the second term in equation (A1) equals $-a^2\partial (\Theta
\Sigma)/\partial \varpi$ with $\Theta$ given by the first relation of
equation (\ref{closure}) rather than by the expression $\Theta =
(3+\lambda^2)/(1+\lambda^2)$ from the analysis of Shu \& Li (1997).
The difference (2 versus 3 in the sum of the numerator) arises because
the latter authors mistakenly included the contribution of
$(B_\varpi^+)^2/8\pi$ into the computation of the magnetic
``pressure'' force, which duplicates a small piece already included in
the first term of equation (A1).

For later reference, we consider the vertical structure if we make the
assumption that the current density is proportional to the volume
density, i.e.,
$$
{\partial B_\varpi \over \partial z} = B_\varpi^+{2\rho\over \Sigma}.
\eqno(\rm{A8})
$$ 
The substitution of equation (A8) into equation (A5) yields a 
differential equation which we may write as
$$
{1\over 8\pi G}\left({\partial U\over \partial z}\right)^2 + 
{a^2\Sigma \over 2B_\varpi^+}{\partial B_\varpi \over \partial z} 
+ {B_\varpi^2\over 8\pi} = {\pi\over 2} G\Sigma^2 
+ {(B_\varpi^+)^2\over 8\pi}.
\eqno(\rm {A9})
$$
The solutions for equations (A4), (A8), and (A9) read
$$
B_\varpi = B_\varpi^+\tanh (z/z_0), \qquad \rho = 
{\Sigma\over 2 z_0}{\rm sech}^2(z/z_0), \qquad 
-{\partial U\over \partial z} = -2\pi G\Sigma \tanh (z/z_0),
\eqno(\rm{A10})
$$
if we set $B_\varpi^+ = B_z = 2\pi G^{1/2}\Sigma/\lambda$.  Equation
(A10) represents the isopedically magnetized version of Camm (1950)
and Spitzer's (1955) solution for the stratified isothermal disk.  We
note in passing, however, that while equation (A7) is an acceptable
approximation in the disk proper, it must fail on the $z$-axis where
$B_\varpi^+ = 0$ from symmetry considerations, but $B_z$ is not only
nonzero but achieves a maximum value there.

\section{Derivation of the Diffusion Constant} 

In this Appendix, we present a brief discussion of the derivation of
the diffusion coefficient in the magnetic diffusion equation.  We
start with the $ z$ component of the induction equation including
ambipolar diffusion
$$ 
{\partial B_z \over \partial t} + {1 \over \varpi} 
{\partial \over \partial \varpi} \left( \varpi B u \right) = 
{1 \over \varpi} {\partial \over \partial \varpi} \left[ 
{\varpi B_z^2 \over 4 \pi \gamma \rho \rho_i} 
{\partial B_\varpi \over \partial z} \right] \, , \eqno({\rm B}1)
$$
where we have kept only the largest term on the right hand side for a
highly flattened core.  Here, $B_z$ is the $ z$ component of the field and
$B_\varpi$ denotes the $ \varpi$ component.  After multiplying the
equation by $\rho$ and integrating over $z$, we obtain the form
$$
\Sigma \left[ {\partial B_z \over \partial t} + {1 \over \varpi} 
{\partial \over \partial \varpi} \left( \varpi B_z u \right) \right] 
= {1 \over \varpi} {\partial \over \partial \varpi} \left[ 
{\varpi B_z^2 \over 4 \pi \gamma} \int_{-\infty}^\infty 
{dz \over \rho_i} {\partial B_\varpi \over \partial z} \right] 
= {1 \over \varpi} {\partial \over \partial \varpi} \left[ 
{\varpi B_z^2 B_\varpi^+ (2 z_0)^{1/2} \over 
2 \pi \gamma {\cal C} \Sigma^{1/2} } \right] \, , \eqno({\rm B}2)
$$ 
where we have defined 
$$
{1\over {\cal C}} \equiv {\Sigma^{1/2} \over (2z_0)^{1/2}B_\varpi^+} 
\int_0^\infty {dz \over \rho_i} {\partial B_\varpi \over \partial z} 
= \int_0^\infty {{\rm sech}(z/z_0)\over {\cal C}_{\rm local}}{dz\over z_0} \, , 
\eqno({\rm B}3)
$$ 
if we assume equation (A10) to hold and express the ion abundance by
the local relation $\rho_i = {\cal C}_{\rm local} \rho^{1/2}$ (Shu
1992). In the present application, ${\cal C}_{\rm local}$ is a
constant if cosmic-rays provide the dominant source of ionization, but
it quickly climbs to much larger values near the surfaces of molecular
clouds because of the ultraviolet ionization of elements like carbon
(McKee 1989). For ${\cal C}_{\rm local}$ = const, we have ${\cal C} =
(2/\pi){\cal C}_{\rm local}$ = $2.0\times 10^{-16}$ cm$^{-3/2}$
g$^{1/2}$ (Shu 1992).

\section{The Velocity Function is Monotonic} 

In this Appendix, we argue that the reduced and scaled velocity field
$ v$ is a monotonic function of $\xi$ for the regime of interest.
Since $ v$ = 0 at $\xi$ = 0 (the inner boundary condition), the
monotonicity of $ v$ implies that the solution must have a nonzero
velocity at large $\xi$. In physical terms, this finding implies that
starless cores are predicted to have nonzero velocities, even before
the collapse phase begins (as observed). Of course, the numerical
integration of the equations of motion implies nonzero values of
${ v}_\infty$. In this Appendix, however, we analytically show
that this property must always hold.

In order to prove this assertion, at least in the context of the
approximations of this paper, it is sufficient to show that the right
hand side of equation (\ref{dervel}) is never equal to zero except at
the critical point $\xi_\ast$ (where the discriminant $\discrim$
changes sign). First, we define the ancillary function
$$P(\xi) \equiv { \sigma} ({\xi +  v}) \, . \eqno({\rm C}1) $$
The right hand side of equation (\ref{dervel}) will be zero if and
only if $P(\xi) = 1$. Further, we know that $P(\xi) = 1$ at the
critical point. Next we show that $P(\xi)$ is monotonic. 
Differentiating $P$ with respect to $\xi$, we obtain 
$${dP \over d\xi} = (\xi + { v}) {d { \sigma} \over d\xi} + 
{ \sigma} \left( 1 + {d { v} \over d\xi} \right) \, . 
\eqno({\rm C}2) $$
Using the equations of motion (eqs. [\ref{deralpha}] and 
[\ref{dervel}]), this result simplifies to the form 
$${dP \over d\xi} = { -\sigma v \over \xi}  . 
\eqno({\rm C}3) 
$$
Since the density $ \sigma$ and the coordinate $\xi$ are 
always positive, and since we are interested in contracting 
solutions where $ v$ is negative, the right hand side of 
this equation, and hence $dP/d\xi$ is positive. As a result, 
the function $P(\xi)$ is a monotonically increasing function 
of the variable $\xi$. 

Since $P(\xi_\ast)$ = 1 and $P(\xi)$ is monotonic, it follows that $P
< 1$ for all $\xi < \xi_\ast$ and $P > 1$ for all $\xi > \xi_\ast$. It
then follows that the right hand side of equation (\ref{dervel}) is
positive for $\xi < \xi_\ast$ and negative for $\xi > \xi_\ast$.
Since the discriminant has the opposite behavior, $\discrim < 0$ for
$\xi < \xi_\ast$ and $\discrim > 0$ for $\xi > \xi_\ast$, it follows
that $- d { v} / d\xi > 0$ for all values of $\xi$. Thus, 
$ v$ is a monotonic function of $\xi$, as claimed. 

\section{Generalized Approach to Crossing Critical Lines} 

In this Appendix, we record the procedure needed to solve the posed
problem when the force integral (\ref{normforce}) has a general form
written as
$$
{ F} \equiv - \Lambda(\xi) {m(\xi) \over \xi^2} = - \Lambda (\xi) 
{ (\xi + { v}) 2 { \sigma} \over \xi} \, . 
\eqno{\rm(D1)}
$$ 
The first equality is true by definition, i.e., we define the function
$\Lambda(\xi)$ to be the ratio of the true force to that given by the
monopole approximation.  The second equality follows from the
continuity equation. THe monopole approximation corresponds to the 
simplest case $\Lambda (\xi)$ = 1. 

With the introduction of the correction function $\Lambda (\xi)$, we
can find the values of the fluid fields and their derivatives at the
critical points. Specifically, for critical point $\xi_\ast$, we find 
$$
{ \sigma} (\xi_\ast) = {1 \over  \Lambda (\xi_\ast) } 
\qquad {\rm and} \qquad { v} (\xi_\ast) = 1 - \xi_\ast \, . 
\eqno{\rm(D2)}
$$ 
Using the same expansion around the critical point as before 
(see eq. [\ref{critexpand}]), we find the derivatives of the 
fluid fields at the critical point, i.e., 
$$
{ v}_1 = - {1 \over 2} \pm {1 \over 2 \xi_\ast} \left[ 
(\xi_\ast - 1)^2 + 1 - 2 \xi_\ast \lamprime / \Lambda_\ast \right]^{1/2} \, , 
\eqno{\rm(D3)}
$$
and 
$$
{ \sigma}_1 = {1 \over 2 \xi_\ast \Lambda_\ast} 
\left\{ \xi_\ast - 2 + 2 \xi_\ast \lamprime / \Lambda_\ast 
\mp \left[ (\xi_\ast - 1)^2 + 1 - 2 \xi_\ast \lamprime / 
\Lambda_\ast \right]^{1/2} \right\} \, . 
\eqno{\rm(D4)}
$$
In these expressions, $\Lambda_\ast = \Lambda(\xi_\ast)$ 
and $\lamprime = d\Lambda/d\xi (\xi_\ast)$.  

\section{Multipole Approach to Effective Sheet Gravity} 

In this Appendix, we consider the multipole approach to the evaluation
of the force equation (\ref{normforce}) (see, e.g., Li \& Shu 1997),
which we write as
$$
F (\xi) = - {dU\over d\xi}, \qquad {\rm where}\qquad 
U(\xi) \equiv \int_0^\infty {\cal H}_0(\xi,\eta)  
\sigma (\eta) \, \eta d\eta \, , 
\eqno{\rm(E1)}
$$
with ${\cal H}_0$ being the classical Poisson kernel for a
self-gravitating axisymmetric sheet, 
$$ 
{\cal H}_0(\xi,\eta) \equiv -{1\over 2\pi}\oint 
{d\varphi \over\sqrt{\xi^2+\eta^2-2\xi\eta\cos\varphi}} .
\eqno{\rm(E2)}
$$  
We now use the well-known formula for the spectral expansion of the
inverse separation distance between a field point at $\xi$ and a
source point at $\eta$ separated by an angle $\varphi$ (see, e.g., 
eq. 3.41 of Jackson 1962):
$$
(\xi^2+\eta^2-2\xi\eta\cos\varphi)^{-1/2} = 
\sum_{\ell = 0}^\infty {\eta^\ell \over \xi^{\ell+1}}P_\ell 
(\cos\varphi) \qquad {\rm for} \qquad \eta < \xi,
\eqno{\rm(E3a)}
$$
$$
(\xi^2+\eta^2-2\xi\eta\cos\varphi)^{-1/2} = 
\sum_{\ell = 0}^\infty {\xi^\ell \over \eta^{\ell+1}}
P_\ell (\cos\varphi) \qquad {\rm for} \qquad \xi < \eta,
\eqno{\rm(E3b)}
$$
where $P_\ell(\mu)$ are the Legendre polynomials of order $\ell$.

For an axisymmetric surface density distribution $\sigma (\eta)$, 
we now have
$$
U(\xi) = -\sum_{\ell=0}^\infty c_{2\ell}\left[U_{2\ell}^<(\xi) 
\xi^{-(2\ell+1)}+U_{2\ell}^>(\xi) \xi^{2\ell}\right], 
\qquad {\rm where} \qquad c_{2\ell} \equiv 
{1\over 2\pi}\oint P_{2\ell} (\cos\varphi)\, d\varphi,
\eqno{\rm(E4a)}
$$
and
$$
U_{2\ell}^< \equiv \int_0^\xi \eta^{2\ell} \sigma(\eta) \, \eta d\eta, 
\qquad U_{2\ell}^> \equiv \int_\xi^\infty \eta^{-(2\ell+1)}
\sigma (\eta) \, \eta d\eta.
\eqno{\rm(E4b)}
$$
Our sum extends over only even values of $\ell$ because the
coefficients $c_\ell$ vanish for odd $\ell$.  The numerical values of
$c_{2\ell}$ are all positive; according to (an equivalent formula by)
Gradshteyn \& Ryzhik (1980, 7.222):
$$
c_{2\ell} = \left[{(2\ell-1)!!\over (2\ell)!!}\right]^2.
$$
Thus, $c_{2\ell}$ = 1, 1/4, 9/64, 25/256, 1225/16384, 3669/65536, etc,
for $2\ell$ = 0, 2, 4, 6, 8, 10, etc, declining slowly only as $\sim
1/2\ell$ for large $\ell$.  We refer to $U_{2\ell}^<$ and $U_{2\ell}^>$ 
as, respectively, the interior (or inner) and the exterior (or outer)
multipole moment of order $2\ell$ with $2\ell = 0, 2,$ etc,
corresponding to the monopole, quadrupole, etc.  In any case, if we
now carry out the differentiation indicated in equation (E1), we get
$$
F(\xi) = \sum_{\ell=0}^\infty c_{2\ell} \left[-(2\ell+1)U_{2\ell}^<(\xi) 
\xi^{-(2\ell+2)}+{2\ell} U_{2\ell}^>(\xi)\xi^{2\ell -1}\right],
\eqno{\rm(E5)}
$$
where we have used the fact that each multipole order cancels in pairs
if we differentiate the moments rather than the powers of $\xi$.  Note
that only the interior monopole moment term $\propto U_0^<$ survives
this differentiation because $2\ell U_{2\ell}^>$ equals zero when
$\ell = 0$.  This well-known result is fortunate since $U_0^>$ is
formally logarithmically divergent if $\sigma(\eta) \rightarrow
A/\eta$ at large $\eta$.

If $\sigma(\eta) = A/\eta$ for all $\eta$ from 0 to $\infty$, then
$U_{2\ell}^< = \xi^{2\ell+1}/(\ell+1)$ and $U_{\ell}^> =
\xi^{-2\ell}/2\ell$ (except for $\ell = 0$), and all multipoles cancel
in pairs in $F$ except for $\ell = 0$.  As is well-kown, the radial
force field for a perfect singular isothermal disk (SID) is given
solely by the interior monopole.  When $\sigma(\eta)$ departs from the
ideal SID state, say by becoming a constant $\sigma_0$ in the central
regions, then $U_{2\ell}^< \approx \sigma_0 \xi^{2\ell+2}/(2\ell+2)$
for small $\xi$, so each interior multipole contributes a constant
term to $F(\xi)$ at small $\xi$.  But the exterior multipole moments
will now contribute terms that are larger, and opposite in sign, to
their interior multipole counterparts.  Thus, the effect of including
multipoles {\it reduces} the inward force of $F(\xi)$ at small $\xi$
relative to the monopole contribution for given $A$.

Computing $F(\xi)$ by equation (\ref{fullforce}) is equivalent to
summing the infinite set of multipole contributions.  In either case,
we can compute the correction function $\Lambda (\xi)$ of Appendix D as
$$
\Lambda(\xi) = {-F\xi^2\over m} , 
\eqno{\rm(E9)}
$$
where $m \equiv U_0^<$.  If the correction function $\Lambda (\xi)$
were known, then one could find the solution using the same procedure
as before (in \S 4 for the monopole solution): Guess the value of the
critical point, move inward from the working estimate of $\xi_\ast$
using the results of \S 6.1, and integrate inward to the origin. Then
adjust the value of the critical point and iterate until the inner
boundary conditions are satisfied. After finding the critical point,
one further integration of the equations of motion (both inward to the
origin and outward to large $\xi$) then determines the solution. In
this case, however, we do not know the function $\Lambda (\xi)$ and
its form depends on the solution for ${ \sigma} (\xi)$ that we are
trying to find.  As a result, we must use another iterative scheme: We
first estimate (guess) the form of the function $\Lambda (\xi)$, and
then calculate the (approximate) solution according to the previous
procedure. With this approximation to $ \sigma$, we can evaluate the
integrals in equations (E4a) and (E4b) to find a new estimate for the
correction function $\Lambda(\xi)$. We then iterate this procedure
until the solution is obtained.

\section{Full Effective Gravity of Unflattened Core}

In this Appendix, we consider the properties of the full gravity of an
incompletely flattened core.  This can be carried out by replacing the
kernel ${\cal H}_0(\xi,\eta)$ by the weighted-average of the product
of the volume densities at the source and field points of the 3-D
Poisson integral (see eq. A9):
$$
{\cal H}(\xi,\eta) \equiv -{1\over 2\pi}\oint d\varphi 
\int_{-\infty}^{+\infty}{d\zeta\over \zeta_0(\xi)}
\int_{-\infty}^{+\infty} {d\zeta^\prime \over \zeta_0(\eta)}  
{{\rm sech}^2[\zeta/\zeta_0(\xi)]{\rm sech}^2[\zeta/\zeta_0(\eta)]
\over \sqrt{\xi^2+\eta^2-2\xi\eta \cos \varphi +(\zeta-\zeta^\prime)^2}}. 
\eqno({\rm F1})
$$
where
$$
\zeta_0(\xi) = {z_0(\xi)\over \Theta_0^{1/2}a|t|} = 
\left[{2(1-f_0^2)\over 1+2f_0^2}\right]{1\over \sigma(\xi)} .
\eqno(\rm{F2})
$$
It is trivial to show that $\partial {\cal H}(0,\eta)/\partial \xi =
0$.  Physically, an axisymmetric magnetized disk with finite thickness
cannot exert a net radial force at its center if its volume density
and current density are regular there.

Although it is possible to develop a multipole expansion procedure for
equation (F1), the resulting analysis would be quite involved.  For
simplicity, therefore, we are content to adopt the alternative
treatment of \S 5.2 designed to give a physical assessment of the
influence of finite disk thickness in the ``softened monopole''
approximation.

%We have tried alleviating the difficulty by replacing the
%Poisson integral ${\cal H}_0$ by the ``softened'' form,
%$$
%{\cal H}_s \equiv - \oint 
%{d\varphi \over \sqrt{\xi^2+\eta^2-2\xi\eta\cos\varphi+\zeta_0^2}} 
%\qquad {\rm where} \qquad  \zeta_0 \equiv 
%{z_0\over a|t|\sqrt{\Theta_0}} = 
%\left({1-f_0^2\over 1+2f_0^2}\right){1\over \sigma}.
%\eqno{\rm(F2)}
%$$
%The form does works and produces the following approximation for small $\xi$ 
%$$
%F = -{\cal F}_s \xi, \qquad {\rm where} \qquad 
%{\cal F}_s \equiv 5\int_0^\infty {\sigma (\eta) \, d\eta \over \eta^2+\zeta_0^2} ,
%\eqno{\rm(F3)} 
%$$
%where ${\cal F}_s$ is a finite positive expression when $\zeta_0$ is
%evaluated with $\sigma =  \sigma_0$.

%However, when we are not at the origin, we need to specify whether
%$\sigma$ in the expression (F2) for $\zeta_0$ is to be evaluated at
%the radius $\xi$ of the field point, or the radius $\eta$ of the
%source point, or by yet some other algorithm.  We have checked that
%simple algorithms do produce reasonable results, with answers
%qualitatively similar to those produced by the monopole approximation;
%however, in detail the answers that one gets depend on the specific
%algorithm adopted. This state of affairs makes difficult any attempt
%to interpret the results physically.  For these reasons, we refrain
%from tabulating any of our numerical explorations, contenting
%ourselves with the observation that substantially better answers on
%this problem than the completely flattened monopole or the monopole
%plus quadrupole calculations will probably require a full 3-D
%treatment in axial symmetry.

\end{document}